\documentclass[manuscript,screen,final]{acmart}

\usepackage{amsmath,amsfonts}

\usepackage{algorithm,algpseudocode}
\algnewcommand{\LeftComment}[1]{\Statex \(\triangleright\) #1}
\usepackage{xparse}
\usepackage{array}
\usepackage{mathtools}
\usepackage{booktabs}
\usepackage{multirow}
\usepackage[caption=false,font=normalsize,labelfont=sf,textfont=sf]{subfig}
\usepackage{textcomp}
\usepackage{adjustbox}
\usepackage{stfloats}
\usepackage{url}
\usepackage{verbatim}
\usepackage{graphicx}

\usepackage{tikz}
\usepackage{pgfplots}
\usepackage{forest}
\algnewcommand\algorithmicswitch{\textbf{switch}}
\algnewcommand\algorithmiccase{\textbf{case}}
\algnewcommand\algorithmicassert{\textt{assert}}
\algnewcommand\Assert[1]{\State \algorithmicassert(#1)}%
\algdef{SE}[SWITCH]{Switch}{EndSwitch}[1]{\algorithmicswitch\ #1\ \algorithmicdo}{\algorithmicend\ \algorithmicswitch}%
\algdef{SE}[CASE]{Case}{EndCase}[1]{\algorithmiccase\ #1}{\algorithmicend\ \algorithmiccase}%
\algtext*{EndSwitch}%
\algtext*{EndCase}%
\AtBeginDocument{%
  }

\setcopyright{acmlicensed}
\copyrightyear{2018}
\acmYear{2018}
\acmDOI{XXXXXXX.XXXXXXX}
\acmConference[Conference acronym 'XX]{Make sure to enter the correct
  conference title from your rights confirmation email}{June 03--05,
  2018}{Woodstock, NY}
\acmISBN{978-1-4503-XXXX-X/2018/06}

\begin{document}

\title{Lightweight Unified \textsc{Sha-3}/\textsc{Shake} Architecture with a Fault-Resilient State}

\author{Christian Ewert\textsuperscript{1},
Amrit Sharma Poudel\textsuperscript{1},
Mouadh Ayache\textsuperscript{1}\textsuperscript{2},
Andrija Neskovic\textsuperscript{1},
Rainer Buchty\textsuperscript{1},
Mladen Berekovic\textsuperscript{1},
Sebastian Berndt\textsuperscript{3},
and Saleh Mulhem\textsuperscript{1}}
\affiliation{\textit{\\
\textsuperscript{1}Institute of Computer Engineering, Universität zu Lübeck, Lübeck, \country{Germany}} \\
\textit{\textsuperscript{2}Synopsys GmbH, Munich, Germany}\\
\textit{\textsuperscript{3}Department of Electrical Engineering and Computer Science, Technische Hochschule Lübeck, Lübeck, Germany}}
\renewcommand{\shortauthors}{Ewert et al.}

\begin{abstract}
       
Hash functions have become a key part of standard Post-quantum cryptography (PQC) schemes, especially \textsc{Sha-3} and \textsc{Shake}, calling for lightweight implementation. 
A fault-resilient design is always desirable to make the whole PQC system reliable. 
We, therefore, propose a) a unified hash engine supporting \textsc{Sha-3} and \textsc{Shake} that follows a byte-wise in-place partitioning mechanism of the so-called \textsc{Keccak} state, and b) an according fault detection for \textsc{Keccak} state protection exploiting its cube structure by deploying two-dimensional parity checks. It outperforms the state-of-the-art (SoA) regarding area requirements at competitive register-level fault detection by achieving 100\% detection of three and still near 100\% of higher numbers of \textsc{Keccak} state faults.
Unlike SoA solutions, the proposed unified hash engine covers all standard hash configurations. 
Moreover, the introduced multidimensional cross-parity check mechanism achieves a 3.7$\times$ improvement in area overhead, with an overall 4.5$\times$ smaller fault-resilient engine design as demonstrated in ASIC and FPGA implementations. Integrated into a RISC-V environment, the unified hash engine with the integrated fault-resilient mechanism introduced less than 8\,\% area overhead.
Our approach thus provides a robust and lightweight fault-detection solution for protecting hash functions deployed in resource-constrained PQC applications.

\end{abstract}



\maketitle


\section{Introduction}
\label{sec:introduction}

\begin{figure}[b]
    \centering
    \includegraphics[trim={0 0 2cm 0},clip,width=.7\columnwidth]{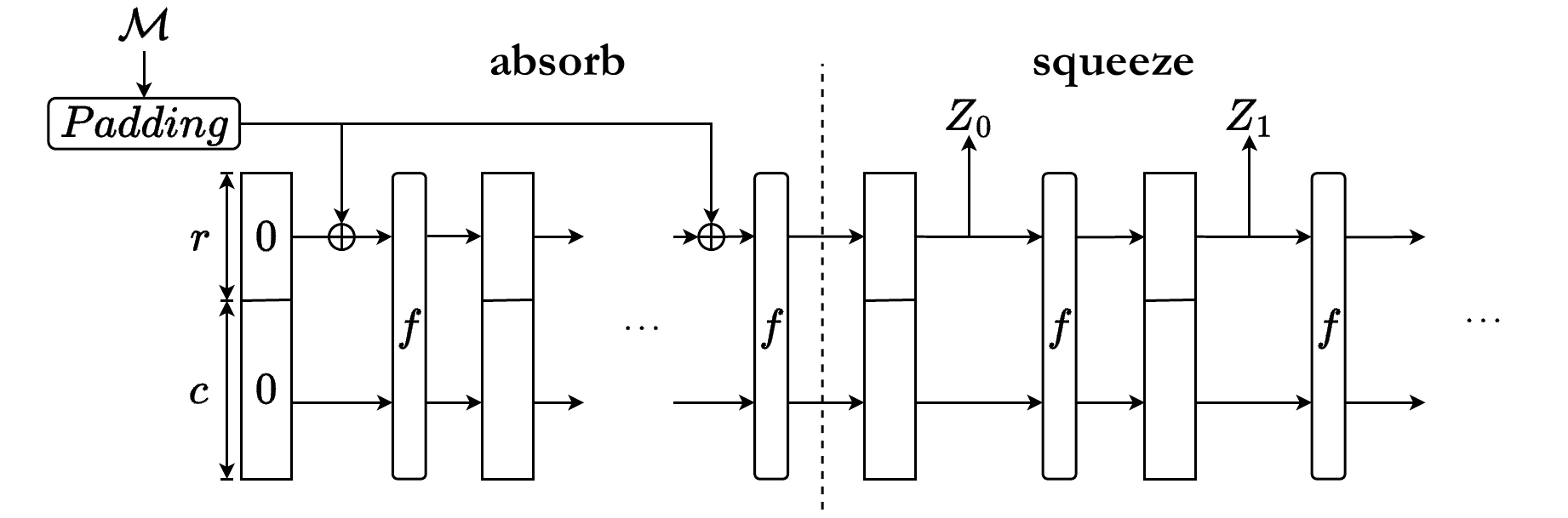}
    \caption{Representation of \textsc{Keccak} Sponge construction \cite{fipsnist}. \textit{f} denotes \\ \textsc{Keccak}-f$[1600]$}
    \label{fig:spongeconstruction}
\end{figure}

Over the years, \textsc{Keccak} (cryptographic hash family) has gained popularity, especially its deployment in the design of \textsc{Sha-3} and \textsc{Shake} standard hash functions. 
The US National Institute of Standards and Technology (NIST) has recently announced post-quantum cryptography (PQC) standard algorithms \cite{PQCDSA24,PQCSLH24,PQCKEM24}; \textsc{Shake} and \textsc{Sha-3} play a crucial role in this standard \cite{PQCSLH24,PQCKEM24}.
For instance, \textsc{Shake} is employed in the PQC standard called Stateless Hash-Based Digital Signature \cite{PQCSLH24}, while \textsc{Sha-3} is a part of Module-Lattice-Based Key-Encapsulation Mechanism Standard \cite{PQCKEM24}. 
Also other PQC algorithms deploy a \textsc{Keccak}-based hash function in their schemes, such as NTRUEncrypt \cite{howgrave2005choosing}, Rainbow \cite{bernstein2017post}, or \textsc{Sphincs}\textsuperscript{+} \cite{bernstein2019sphincs+}. 
Note that in the following sections, we generally use \textsc{Keccak} as illustrated in Fig.~\ref{fig:spongeconstruction}. 
The bit sizes of the so-called rate $r$ and capacity $c$ with the message (input) padding mechanism determine if \textsc{Keccak} configuration serves as \textsc{Sha-3} or \textsc{Shake}.
Therefore, we only explicitly use \textsc{Sha-3} and \textsc{Shake} when needed. 

\subsection{Related Work}
\label{subsec:Related}

\subsubsection{Unified Cryptographic Engine}
It has always been desirable to design a unified unit or engine of cryptographic primitives \cite{mishra2020high} \cite{khan2007design}. 
Such a unified engine can serve several modes and has become very important for PQC \cite{aikata2022unified}.  
For instance, HMAC-Hash unit was proposed in \cite{khan2007design} covering six cryptographic hashes, namely: MD5, SHA-1, RIPEMD-160, HMAC-MD5, HMAC-SHA-1, and HMAC-RIPEMD-160 implemented in parallel.
The HMAC-Hash unit design deploys some bit controls to select whether the HMAC algorithm or hash function is to be the only one executed. 
In\cite{at2017low}, a compact 8-bit unified coprocessor for AES and Grøstl hash function was designed for resource-constrained embedded systems. 
This coprocessor considers three AES modes with 128-, 192-, and 256-bit encryption keys and two Grøstl hash modes with  256- and 512-bit message digests. 
Similarly, the three AES modes were unified with the ECHO hash function as one coprocessor \cite{beuchat2011low}.
A unified AES and SHA-3 engine was proposed in \cite{khalid2020resource}. 
This engine relies on a unified XOR section that executes key whitening in AES and SHA-3 transformations.
\begin{table}[t]
\centering
\caption{State of the Art Overview}
\label{tab:soa_overview_reduced}
\begin{tabular}{cccc}
\toprule
\textbf{Paper}           & \textbf{\begin{tabular}[c]{@{}c@{}}Protection\\ Level\end{tabular}} & \textbf{\begin{tabular}[c]{@{}c@{}}Protection\\ Mechanism\end{tabular}} & \textbf{\begin{tabular}[c]{@{}c@{}}Hardware\\ Overhead\end{tabular}} \\ \midrule
\cite{kahri2017}         & Logic                                                               & Fault Detection                                                        & High                                                                 \\ 
\cite{bayatsarmadi2014}  & Logic                                                               & Error Detection                                                        & Mid                                                                  \\ 
\cite{mestiri2021}       & Logic                                                               & Error Detection                                                        & High                                                                 \\ 
\cite{mestiri2023}       & Logic                                                               & Error Detection                                                        & High                                                                 \\ 
\cite{purnal2019}        & Logic                                                               & Fault Detection                                                        & High                                                                 \\ 
\cite{luo2016}           & Logic                                                               & Error Detection                                                        & Mid - High                                                           \\ 
\cite{torresalvardo2022} & \begin{tabular}[c]{@{}c@{}}Register \\ \& Logic\end{tabular}        & Error Correction                                                       & High                                                                 \\ 
\cite{gavrilan_2024}     & Logic                                                               & Error Detection                                                        & High                                                                 \\ 
This  work          & Register                                                            & Fault Detection                                                        & Low                                                                  \\ \bottomrule
\end{tabular}
\end{table}
In the content of PQC, several unified hash architectures for \textsc{Sha} and \textsc{Shake} have been proposed to accelerate different PQC algorithms. 
The key idea of the implementation relies on in-place partitioning of \textsc{Keccak} state to serve several hash modes.
The in-place partitioning mechanism is performed based on 64-bit data word size. 
In \cite{roy2020high}, a unified hash architecture for \textsc{Sha-3-256}, \textsc{Sha-3-512}, and \textsc{Shake-128} was proposed to be a part of the SABER PQC algorithm.  
Similarly, a unified hash architecture for \textsc{Sha-3-512}, \textsc{Shake-128}, and \textsc{Shake-256} was proposed in \cite{xi2023low} to accelerate Dilithium and Kyber PQC algorithms on a field-programmable gate array (FPGA) board.
This unified implementation also uses a 64-bit data word in-place partitioning of \textsc{Keccak} state. 
Therefore, such a wide in-place partitioning mechanism makes SA unified architectures very costly in hardware and unsuitable for resource-constrained devices. 

Furthermore, a reliable and secure implementation of a unified engine for cryptographic primitives has emerged as a new design prescriptive. 
In the case of fault injection, a fault-resilient unified engine for cryptographic primitives is highly desirable.
However, few unified engines consider this design prescriptive (e.g., \cite{zhao2012unified}).
Thus, there is still a huge room for innovation and improvement. 
 
\subsubsection{Fault Resilient \textsc{Sha-3} and \textsc{Shake} Engine}
In the following, we quickly review the state-of-the-art (SoA) fault-resilient mechanisms of \textsc{Sha-3} and \textsc{Shake} engines.
In \cite{rama2024trustworthy}, faults are divided into two classes based on their causes: (i) intentional malicious faults covered in the security domain and (ii) random faults covered in the reliability domain. 
A fault-resilient design increases the overall dependability of the unified engine. 
However, fault-resilient mechanisms can protect the desired engine entirely or partially based on the targeted application \cite{rama2024trustworthy}. 
Here, we focus on the designs of fault resilient \textsc{Sha-3} and \textsc{Shake}, that can handle the faults and their propagation independently from the causes. 
For example, Differential Fault Analysis (DFA) \cite{biham_dfa, bagheri_dfa, luo_dfa} was performed on complete 24 \textsc{Keccak} rounds. 
Such an analysis shows that DFA can potentially reveal the whole \textsc{Keccak} state.  
DFA is known to be a powerful method to compromise many cryptographic primitives like \textsc{Des}, \textsc{3-Des}, \textsc{Idea} \cite{biham_dfa}, \textsc{Aes} \cite{Giraud2012} and \textsc{Rsa} \cite{Berzati2012}. 
In \cite{bagheri_dfa}, DFA was introduced as a random single-bit fault injected into the internal state of \textsc{Keccak} at the beginning of the penultimate (22\textsuperscript{nd}) round.
In \cite{luo_dfa}, the analysis of DFA on \textsc{Sha-3} described in \cite{bagheri_dfa} was extended to DFA on a byte-granular level in a similar fashion to \cite{bagheri_dfa}. 
The injected faults randomly alter a state byte, leading to faulty and non-faulty hashes of the same message.  
In contrast to the single-bit fault model, it applies to all four modes of \textsc{Sha-3}.
This method can effectively identify the injected single-byte fault and recover the whole internal state of \textsc{Keccak}.
Therefore, several fault-resilient approaches have been proposed for \textsc{Sha-3} and \textsc{Shake} engines. 

In the SoA, several works have been carried out and proposed a fault resilient \textsc{Keccak}.
Table~\ref{tab:soa_overview_reduced} shows an overview of SoA fault resilient \textsc{Keccak} countermeasures.
We mainly categorize in \textit{fault detection}, \textit{error detection}, and \textit{error correction}.
All of the mentioned detection approaches focus on the protection of \textsc{Keccak} layers at the logic level \cite{kahri2017,bayatsarmadi2014,mestiri2021,purnal2019,luo2016,gavrilan_2024}.
Only one error-correction approach also considers register-level protection \cite{torresalvardo2022}, providing a holistic protection mechanism for \textsc{Keccak}.
The protection mechanisms mentioned at the logic level may give strong security guarantees but exhibit mid- to high hardware overhead, making them unsuitable for resource-constrained devices.

\subsection{Motivation \& Contributions}
\label{subsec:paper_idea}
\begin{figure}[t]
    \centerline{\includegraphics[width=.45\columnwidth]{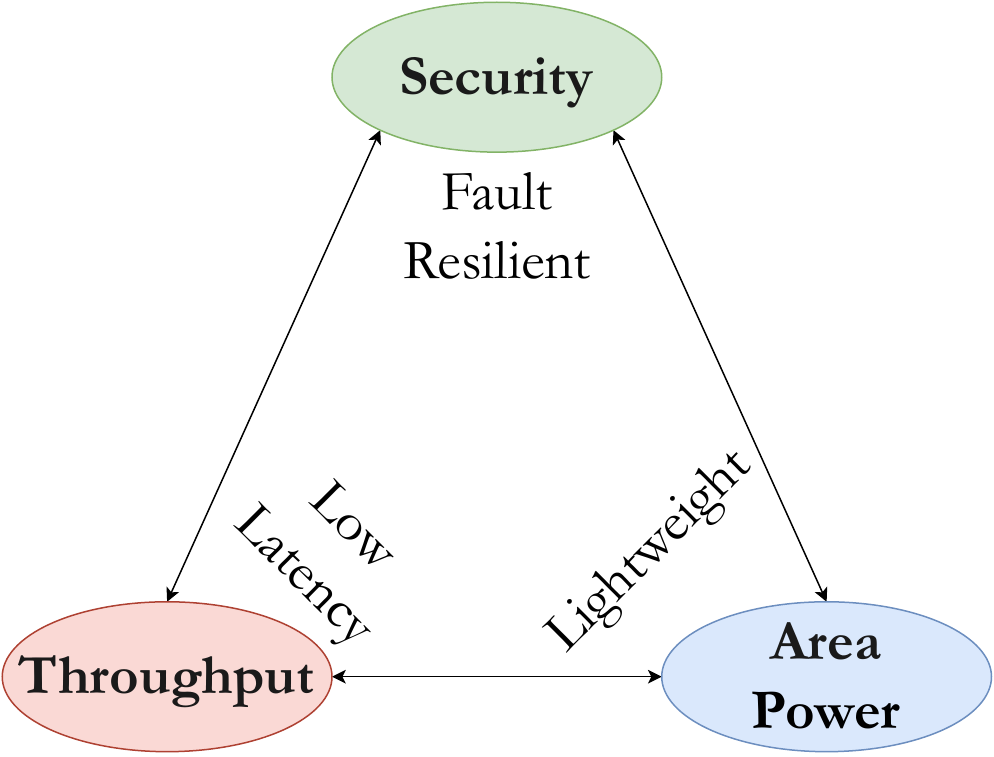}}
    \caption{Trade-offs in Cryptographic Hardware Design}
    \label{fig:trade-offsCrypto}
\end{figure}

Besides the classical digital signatures, data integrity, and several forms of authentication, \textsc{Sha-3} and \textsc{Shake} are deployed in several parts of PQC algorithms \cite{PQCSLH24,PQCKEM24}.   
Therefore, a unified architecture of \textsc{Sha-3} and \textsc{Shake} can simultaneously serve in modern and post-cryptographic systems. 
This fact first motivates us to design such a unified engine.
The SoA requirements for the design vary based on the application domain. 
Therefore, our second motivation is to follow the design principle of lightweight cryptography \cite{LWC2018} and choose the design objectives that focus on the hardware cost, low latency, and fault resiliency as shown in Fig.~\ref{fig:trade-offsCrypto}. 
The paper's motivations can be summarized as follows: 
\begin{itemize} 
    \item A unified architecture of \textsc{Sha-3} and \textsc{Shake} covering all their standard modes \cite{fipsnist}: \textsc{Sha-3-224}, \textsc{Sha-3-256}, \textsc{Sha-3-384}, \textsc{Sha-3-512}, \textsc{Shake128}, and \textsc{Shake-256}.
    \item The proposed unified engine is fault-resilient at minimum hardware overhead. 
    In particular, the proposed fault-detection mechanism meets the requirements of lightweight cryptography:
    We focus on register-level protection that provides a low-overhead fault countermeasure of \textsc{Keccak}.
\end{itemize}

\textsc{Sha-3} and \textsc{Shake} have a huge \textsc{Keccak} state of $1600$ bits.  
This state has a dedicated configuration of its parameters (rate $r$ and capacity $c$) to serve as \textsc{Sha-3-224}, \textsc{Sha-3-256}, \textsc{Sha-3-384}, \textsc{Sha-3-512}, \textsc{Shake-128}, and \textsc{Shake-256}. 
This is exactly the main design challenge of a unified \textsc{Sha-3}/\textsc{Shake} engine.
To cover all standard hash modes, partitioning into two parts with variable sizes according to the hash modes is required. 
Moreover, it makes the design of the fault-resilient state even more complex.   
Addressing this challenge, our contributions to this work are listed as follows:
\begin{enumerate}
    \item We introduce a new in-place state-partitioning mechanism. 
    The proposed mechanism is a byte-wise state partitioning. 
    Thus, it is flexible, lightweight, and can serve in all required hash standard modes, namely: \textsc{Sha-3-224}, \textsc{Sha-3-256}, \textsc{Sha-3-384}, \textsc{Sha-3-512}, \textsc{Shake-128}, and \textsc{Shake-256}. 
    This allows building a unified \textsc{Sha-3}/\textsc{Shake} engine that meets the requirements of lightweight cryptography, specifically regarding area and power. 
    \item For reliable and secure implementation, we propose a new low-overhead fault-resilient mechanism for \textsc{Keccak} state.  
    It is a multi-dimensional cross-parity check, exploiting the cube structure of the \textsc{Keccak} state. 
    First, we present a one-dimensional (column) parity check. 
    Second, the proposed mechanism is extended by two-dimensional parity: one-dimensional parity for the lane and one more for the sheet to realize a three-dimensional cross-parity, detecting up to three faulty bits in the \textsc{Keccak} state.
    \item Since RISC-V is a key player for future embedded system-on-chip and microcontroller design, we integrate the resulting fault-resilient unified hash engine into a 32-bit RISC-V-based microcontroller and implement it using a 45\,nm standard library and in a Virtex-7 FPGA.
    The hardware implementation results show that the integration of the proposed unified \textsc{Sha-3}/\textsc{Shake} engine into a 32-bit RISC-V-based microcontroller results in a suitable approach for resource-constrained embedded systems applications, especially for PQC at the edge level.
\end{enumerate}

\subsection{Outline}
\label{subsec:outline}
First, the description of \textsc{Sha-3} and \textsc{Shake} in general and \textsc{Keccak} in particular is provided in Section~\ref{sec:preliminaries}. 
The new design of a unified \textsc{Sha-3}/\textsc{Shake} engine architecture is proposed in Section~\ref{sec:unified_hash_engine}.
Subsequently, the new fault-resilient mechanism of \textsc{Keccak} state is illustrated in Section~\ref{sec:lightweight_fault_detection_mechanism}.
The corresponding implementation results of the unified \textsc{Sha-3}/\textsc{Shake} engine and the fault--resilient mechanism are presented in Section~\ref{sec:results} followed by the findings of the paper summarized in Section~\ref{sec:conclusion}.
\section{BACKGROUND}
\label{sec:preliminaries}
This section highlights the definitions and notation used in this work. 
It then describes the hash functions SHA-3 and SHAKE in general and, in particular, KECCAK as the core component of the overlying hash functions.

\subsection{Notation}
\label{subsec:notation}
We use the following notation:
We denote a message consisting of $n$ bytes of data as $\mathcal{M}$ = [$m_1, m_2, \ldots, m_n$], where $m_i$ specifies the $i^{th}$ message byte.
Similarly, the generated hash value comprising $m$ bytes of data is described as $\mathcal{H}$ = [$h_1, h_2, \ldots, h_t$], where $h_k$ specifies the $k^{th}$ hash byte.
Furthermore, we denote \boldmath$S$\unboldmath\, as an array representing the bit-wise \textsc{Keccak} state, and $S[x,y,z]$ as a corresponding bit, where every other array throughout the paper is described similarly.
To prohibit the disclosure of the faulted hash, we denote signalizing a detected fault via the signal \textit{error}.

\subsection{S{\footnotesize HA-3} and S{\footnotesize HAKE}}
\label{subsec:sha3_shake}

In 2012, NIST declared \textsc{Keccak} \cite{guidocsf, guidosubmission} the winner of the \textsc{Sha-3} competition, which was initiated in 2007. This competition aimed to create a new cryptographic hash function to complement the existing \textsc{Sha-2} family. Consequently, NIST published FIPS202 \cite{fipsnist} in 2015, which standardized the hash-function family \textsc{Sha-3}. In contrast to its predecessors, \textsc{Keccak} is based on the novel sponge construction concept as described in \cite{guidocsf}. \textsc{Keccak} has since then found applications in many other cryptographic primitives like \textsc{SpongeWrap} \cite{guidocsf} and \textsc{Kmac} \cite{kelsey2016sha}, \textsc{Spongent} \cite{bogdanov2011spongent}. 

\textsc{Sha-3} comprises four distinct hash functions, each distinguished by the length of the resulting digest they produce. These variants are known as \textsc{Sha3-224}, \textsc{Sha3-256}, \textsc{Sha3-384}, and \textsc{Sha3-512}, where the suffix following the hyphen signifies the size of the output digest. In addition, this family also consists of two extendable-output functions (\textsc{Xof}s), named \textsc{Shake-128} and \textsc{Shake-256} \cite{fipsnist}. For these functions, the output length can be chosen arbitrarily to meet the requirements of individual applications \cite{fipsnist}. The proposed term \textsc{Shake} combines the term \textit{Secure Hash Algorithm} with \textsc{Keccak} \cite{fipsnist}. They all share the common underlying Sponge construction, thus making them examples of sponge functions.

\subsection{K\footnotesize ECCAK}
\label{subsec:keccak}
As the foundation of all \textsc{Sha-3} and \textsc{Shake} hashes, \textsc{Keccak} relies on the sponge construction as shown in Fig.~\ref{fig:spongeconstruction}. The sponge construction works on input data with variable length and arbitrary output length, as proposed by Bertoni et al. \cite{guidocsf, fipsnist}.

\begin{figure}[t]
    \centering
    \includegraphics[width=.55\columnwidth]{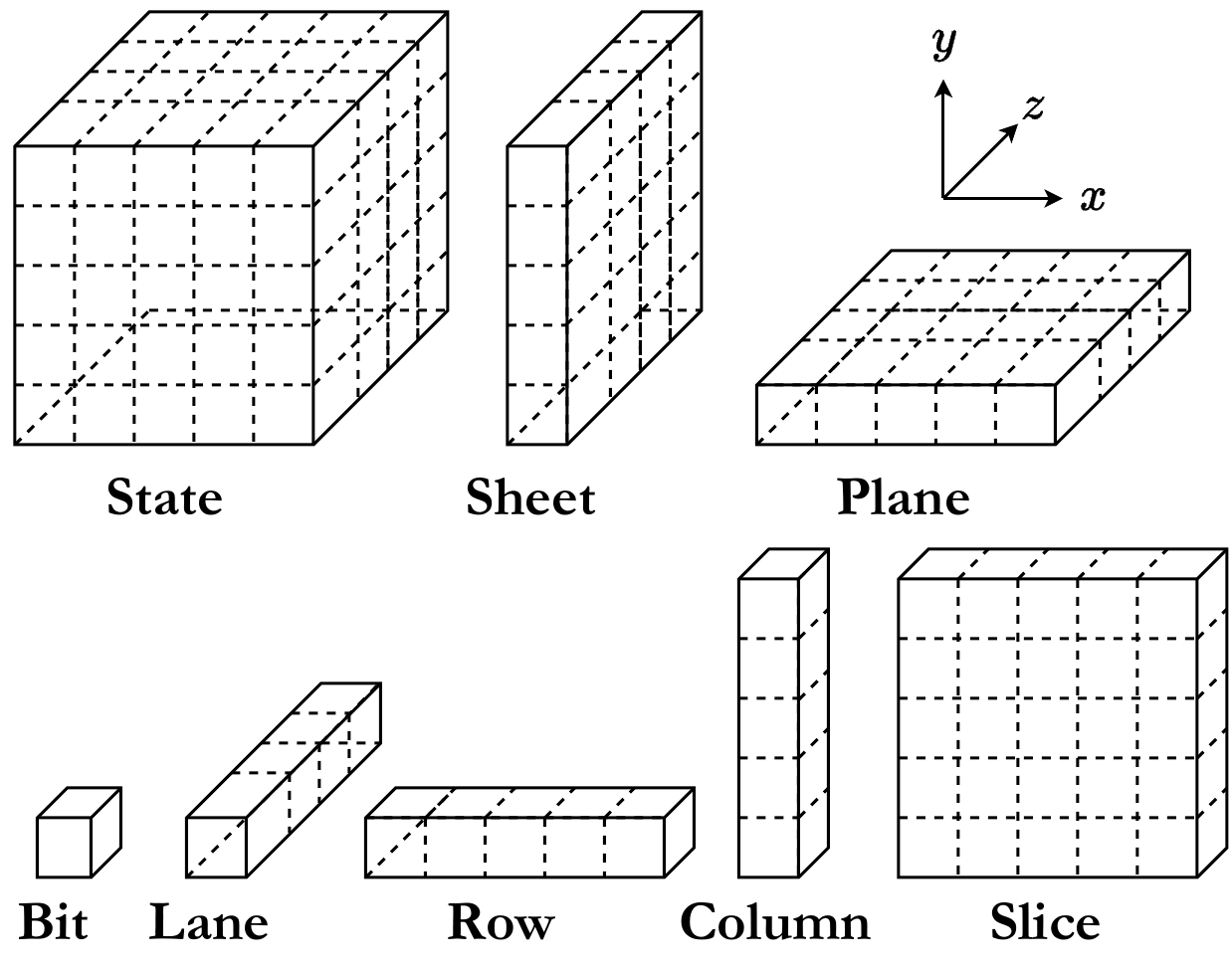}
    \caption{Naming convention for the subelements of a \textsc{Keccak} state \cite{fipsnist}.}
    \label{fig:keccakstate}
\end{figure}

The sponge construction consists of a three-dimensional state $b = x \times y \times 2^{l} = 5 \times 5 \times 64 = 1600$ for the standardized \textsc{Sha-3} and \textsc{Shake}, where $l$ is defined as $l = \log_{2}(b/25)$.
The state (block) consists of sub-elements noted as \textit{column}, \textit{lane}, \textit{row}, \textit{sheet}, \textit{slice} and \textit{plane} as shown in Fig.~\ref{fig:keccakstate}.
 \\
A \textit{sheet} consists of 64 \textit{columns} in $z$- or 5 \textit{lanes} in $y$-direction, respectively.
A \textit{slice} is built out of 5 \textit{columns} in $x$- or 5 \textit{rows} in the $y$-direction.
A plane is described as an $x$/$z$-direction layer consisting of $5 \times 64$ \textit{bits}, 64 \textit{rows}, or 5 \textit{lanes}, respectively.
To build the described elements, column and row contain 5 bits each, and a lane comprises 64 bits.
The state $b$ is split by the \textsc{Sha-3} and \textsc{Shake} mode-dependent rate $r$ and capacity $c$ in two consecutive parts, such that $b = r+c$.
For a hash generation, the input message is padded based on \textsc{Sha-3} or \textsc{Shake} to multiple blocks with $r$-size as shown in Algorithm ~\ref{alg:padding}, resulting in a string $P$ such that $m+len(P)$ is a positive multiple of $r$.

\begin{algorithm}[h]
    \caption{Pad \textsc{Sha-3} / \textsc{Shake} \cite{fipsnist}}
    \label{alg:padding}
    \begin{flushleft}
    \hspace*{\algorithmicindent} \textbf{Input:} Rate size $r$, Message size $m$, Hash Mode $mode$ \\
    \hspace*{\algorithmicindent} \textbf{Output:} String $P$ such that $m+len(P)$ is a positive multiple of $r$.
    \end{flushleft}
    \begin{algorithmic}[1]

        \If{mode == \textsc{SHA-3}}
            \State $\phantom{P}\mathllap{j} \gets (-m-4)$ mod $r$
            \State $P \gets 01 || 1 || 0^{j} || 1$
        \ElsIf{mode == \textsc{SHAKE}}
            \State $\phantom{P}\mathllap{j} \gets (-m-6)$ mod $r$
            \State $P \gets 1111 ||1 || 0^{j} || 1$
        \EndIf
        
        \State \Return $P$
        
    \end{algorithmic}
\end{algorithm}

Then, it is \textsc{Xor}'ed with the rate $r$ as illustrated in Fig.~\ref{fig:spongeconstruction} before.
For every $r$-sized block and the current capacity $c$, the \textsc{Keccak}-$f[b]$ permutation for $n_{r} = 12+2l = 24$ rounds is performed as shown in Algorithm~\ref{alg:keccak}:

\begin{algorithm}[h]
    \caption{\textsc{Keccak}-f[1600] \cite{fipsnist}}
    \begin{flushleft}
    \hspace*{\algorithmicindent} \textbf{Input:} \textsc{Keccak} State Array \boldmath$S$\unboldmath, Round Constant \boldmath$rc$\unboldmath \\
     \hspace*{\algorithmicindent} \textbf{Output:} \textsc{Keccak} State Array \boldmath$S'$\unboldmath
     \end{flushleft}
    \begin{algorithmic}[1]
    \ForAll{$0 \leq i < 24$}
        \ForAll{triples $(x,y,z)$ such that $0 \leq x < 5, 0 \leq y < 5$ and $0 \leq z < 64$ }
            \State $\phantom{S'[x,y,z]}\mathllap{\theta[x,z]} \gets S[x,y,z] \oplus (\oplus_{y=0}^{4} S[x-1,y,z]) \oplus (\oplus_{y=0}^{4} S[x+1,y,z-1])$
            \State $\phantom{S'[x,y,z]}\mathllap{\rho[x,y,z]} \gets \theta[(x+3y) \text{ mod } 5,x,z]$
            \State $\phantom{S'[x,y,z]}\mathllap{\pi[x,y,z]} \gets \rho[(x+3y) \text{ mod } 5,x,z]$ 
            \State $\phantom{S'[x,y,z]}\mathllap{\chi[x,y,z]} \gets \pi[x,y,z] \oplus ((\pi[(x+1) \text{ mod } 5,y,z] \oplus 1) \cdot \pi[(x+2) \text{ mod 5}, y, z]) $ 
            \State $\phantom{S'[x,y,z]}\mathllap{\iota[0,0,z]} \gets \chi[0,0,z] \oplus rc[i,z]$
            \State $S'[x,y,z] \gets \chi[x,y,z], S'[0,0,z] = \iota[0,0,z]$
        \EndFor
    \EndFor
    \State {Return \boldmath$S'$\unboldmath}
    \end{algorithmic}
    \label{alg:keccak}
\end{algorithm}

In the $\theta$ layer, the sum of every column is generated and interlocked with a corresponding column of the state. In the layers $\rho$ and $\pi$, bit-shuffling in every \textit{sheet} and \textit{slice} is performed, followed by the non-linear layer $\chi$. In $\iota$, the round-constant $rc(i)$ of the corresponding round $i$ is added to the mid-lane.
After the padded massage is \textit{absorbed}, the digest is \textit{squeezed} by taking the data from the \textit{rate} and refresh the state by \textsc{Keccak}-$f[1600]$ if the digest size is greater than the size of the rate as shown in Fig.~\ref{fig:spongeconstruction}.

\section{Unified \textsc{SHA-3}/\textsc{Shake} Architecture}
\label{sec:unified_hash_engine}

In this section, we introduce a unified low-cost hardware architecture for both \textsc{Sha-3} and \textsc{Shake} standard hash functions.
In particular, we present the design of a byte-wise in-place state-partitioning mechanism. 
This lightweight partitioning mechanism allows the building of a unified \textsc{Sha-3}/\textsc{Shake} engine for all required hash modes that meet the requirements of lightweight cryptography, specifically regarding area and power.

Since both hash functions have \textsc{Keccak} core as a key building block, the proposed unified engine of \textsc{Sha-3} and \textsc{Shake} can share and jointly use a \textsc{Keccak} core.   
The design decisions are made to achieve low power, low area consumption, and high throughput, targeting embedded and IoT applications.
For this purpose, the input message $m_i$ and the hash output $h_k$ work on a byte-granular data resolution.

\begin{figure}[t]
\centerline{\includegraphics[trim={1.5cm 3.8cm 1cm 0.5cm},clip,width=.8\columnwidth]{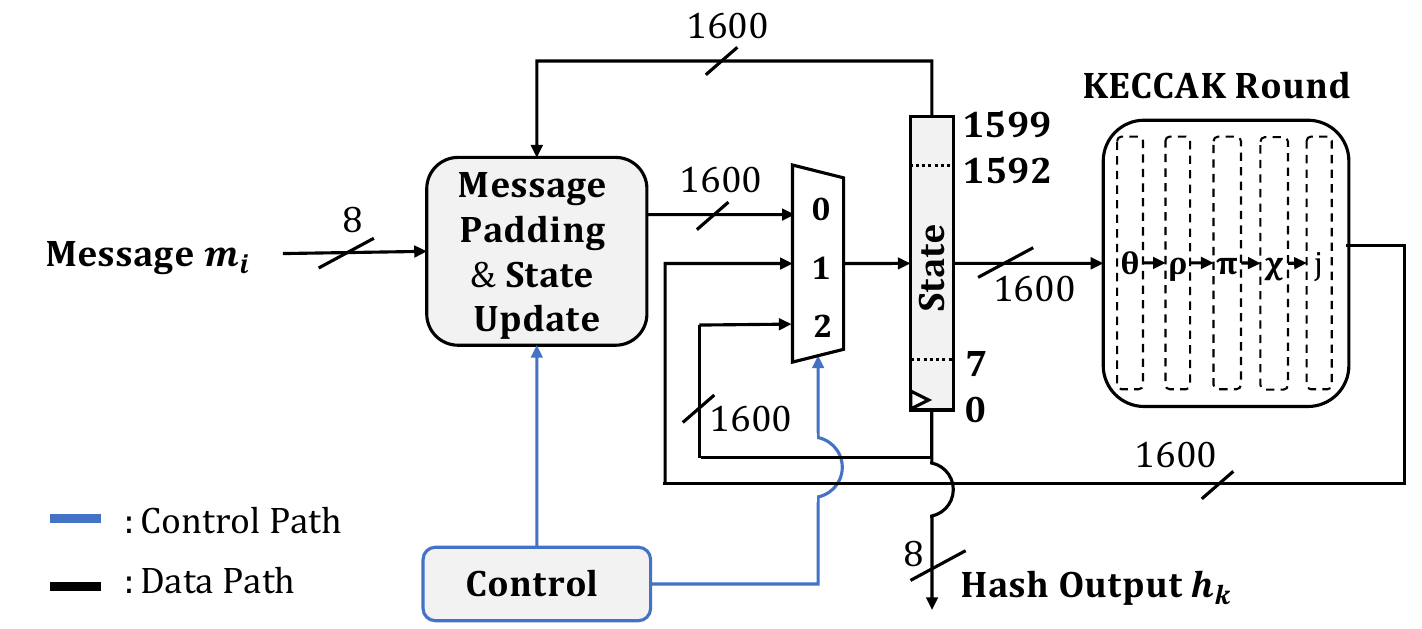}}
\caption{\textsc{Sha-3}/\textsc{Shake} Unified Engine Design}
\label{fig:sha3_shake_basic_engine}
\end{figure}
 
\begin{figure}[b]
    \centerline{\includegraphics[trim={0cm 3.9cm 2.1cm 0.5cm},clip,width=.85\columnwidth]{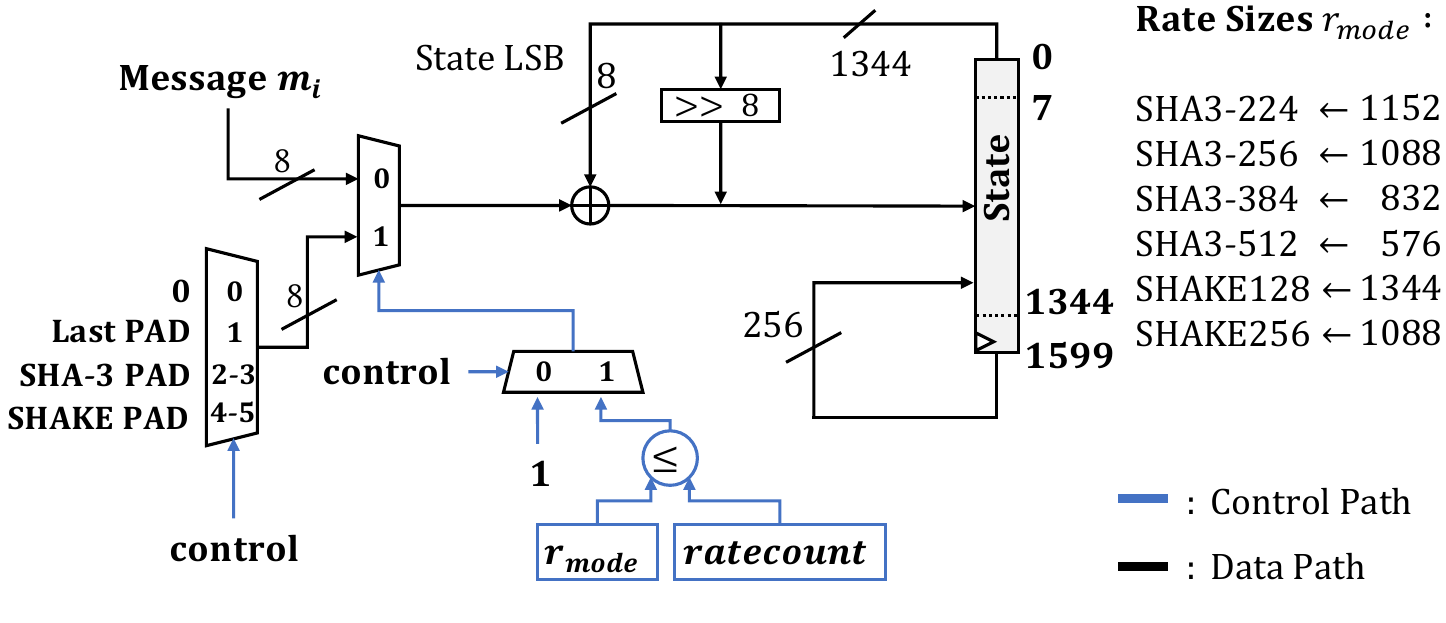}}
    \caption{\textsc{Keccak} State Update by (Padded) Message}
    \label{fig:sha3_shake_basic_engine_state_update}
\end{figure}

Fig.~\ref{fig:sha3_shake_basic_engine} depicts the proposed unified hash engine. 
It comprises a \textsc{Keccak} core, a padding and update unit, a control unit, and a register holding the \textsc{Keccak} state.
Considering the \textsc{Sha-3} and \textsc{Shake} standards, the size of the \textsc{Keccak} state \boldmath$S$\unboldmath\, is 1600 bits. 
This huge state contributes a large area to the overall engine design.
The proposed engine utilizes in-place data processing (byte-wise) to minimize the impact of the corresponding state register i.e., the state is byte-wise logically partitioned and has a special padding and update unit to cover all \textsc{Sha-3} and \textsc{Shake} standard hash modes, namely \cite{fipsnist}: \textsc{Sha-3-224}, \textsc{Sha-3-256}, \textsc{Sha-3-384}, \textsc{Sha-3-512}, \textsc{Shake-128}, and \textsc{Shake-256} as shown in Fig.~\ref{fig:sha3_shake_basic_engine_state_update}.

We consider two different rate sizes to support all the hash modes.
The first rate size is denoted as $r_{sr} = 1344$, which is pre-determined by the \textsc{Shake-128} rate size and the largest of all \textsc{Sha-3} and \textsc{Shake} hash modes.
Only the part $S$[$1343$\,:\,$0$] of the \textsc{Keccak} state specified by $r_{sr}$ is updated and configured as a shift register to enable in-place data processing and keeps the state update delay as low as possible.
The state's capacity part $S$[$1599$\,:\,$1344$] consists of $c_{sr} = 256$ bits and is configured only to be updated by the output of \textsc{Keccak}.
The second rate size $r_{mode}$ is determined using the hash mode. 
If a hash mode other than \textsc{Shake-128} is selected, the mode-specific rate $r_{mode}$ is smaller than $r_{sr}$, otherwise $r_{mode}$ is equal to $r_{sr}$.
The padding and update unit performs the state update in place by \textsc{Xor}'ing the message byte $m_i$ with the state's least significant byte (LSB) and appending it to the most significant byte (MSB) of the rate as 
\begin{equation}
   S'[1343\text{\,:\,}0]  \leftarrow (m_i \oplus S'[7\text{\,:\,}0]) \mathbin\Vert (S[1343\text{\,:\,}0]  \gg 8)\ .
   \label{eq:rate_update}
\end{equation}
A \textit{ratecount} increases with every consumed message byte. 
When $ratecount = r_{mode}$, the mode-specific rate $r$ update is finished. 
To keep the state consistent, the remaining $r_{sr} - r_{mode}$ bytes are \textsc{Xor}'ed with zero bytes and appended to the MSB side of the shift register as described before.
This mechanism makes the remaining bytes extend the capacity size $c_{sr}$ for the corresponding hash mode by $r_{sr} - r_{mode}$ bytes.
To perform the padding according to \textsc{Sha-3} and \textsc{Shake} as described in Algorithm~\ref{alg:padding}, a 5-to-1 multiplexer selects the corresponding padding byte, so the state update can be performed as described before.
Further, when the shift register part $S$[$1344$\,:\,$0$] of the \textsc{Keccak} state is updated by the (padded) message and the mode-dependent capacity if needed, the \textsc{Keccak}-$f$ permutation is performed.
To achieve a low-area and low-power design while ensuring sufficiently high performance, the permutation works as a round-based implementation, updating the state without introducing an additional register.
To reduce the computing latency, the \textsc{Keccak}-$f$ permutation rounds can be unrolled for the cost of additional hardware overhead.
The output of a hash byte $h_k$ is performed by selecting the LSB of the \textsc{Keccak} state $S[7:0]$.
To keep the state consistent, the rate update is performed as described before by \textsc{Xor}'ing the hash byte $h_k$ with a zero byte and appending it to the MSB side of the shift register. 
The \textsc{Keccak} state is preserved to perform a state update by \textsc{Keccak} in the case more hash bytes are requested than the rate provides.
This procedure is repeated until all requested hash bytes are transmitted.

\section{Lightweight Fault Detection Mechanism for \textsc{Keccak} State}
\label{sec:lightweight_fault_detection_mechanism}

This section presents the lightweight fault detection (FD) mechanism for \textsc{Keccak}.
The main goal of the proposed FD approach is to protect the \textsc{Keccak} state against faults at minimal hardware cost.
The proposed FD mechanism combines spatial and information redundancy as FD schemes. 
Then, we apply the proposed FD mechanism in the unified hash engine.
We show the FD mechanism's flexibility and integration in two realizations of the unified hash engine, e.g., round-based and unrolled implementations.
To demonstrate, we integrate the resulting fault-resilient unified hash engine into a RISC-V-based microcontroller.  

\subsection{Spatial \& Information Redundancy Scheme for Fault Detection}
\label{subsec:combined_fault_detection_scheme}
\begin{figure}[t]
\centerline{\includegraphics[trim=3cm 1.2cm 8.5cm 1.0cm, clip, width=.5\columnwidth]{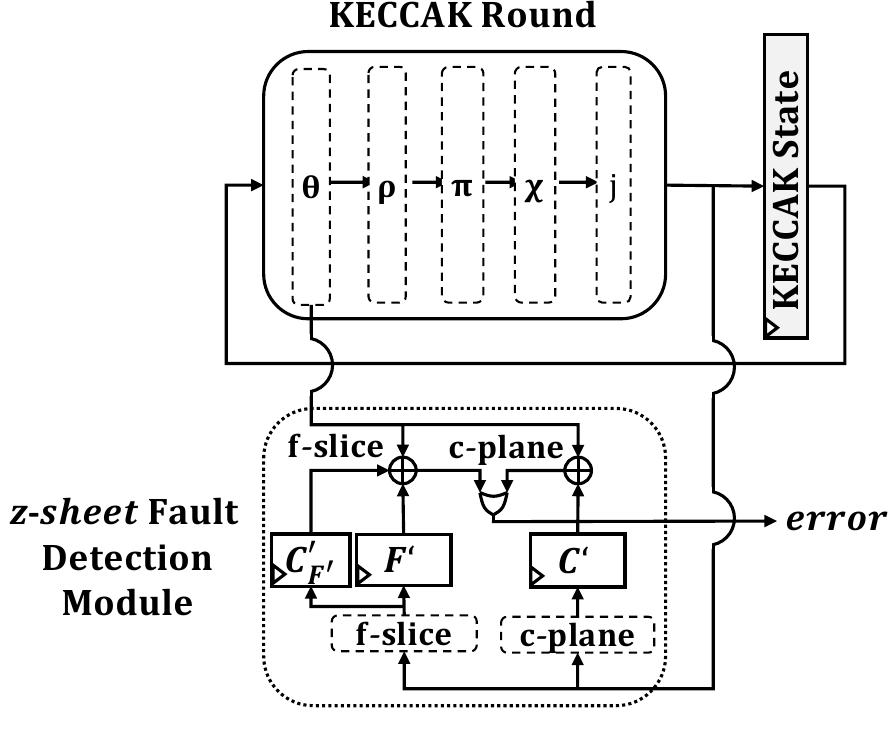}}
\caption{Combined FD Schemes for \textsc{Keccak}}
\label{fig:combined_fault_detection_scheme}
\end{figure}

Fig.~\ref{fig:combined_fault_detection_scheme} shows the proposed FD mechanism as a hardware module placed alongside \textsc{Keccak} realizing spatial redundancy.
The proposed mechanism consists of two main building blocks: \textit{c-plane} and \textit{f-slice} as one-dimensional parity check schemes.
Both one-dimensional parity checks are combined to provide a two-dimensional parity check along a \textit{sheet}, denoted by \textit{z-sheet}.
Therefore, the proposed multi-dimensional parity check mechanism can detect up to three-bit flips.    

\subsubsection{\textit{c-plane}-based Parity Check}
\label{subsec:c_plane_parity_check}
Algorithm~\ref{alg:theta} illustrates the theta layer of \textsc{Keccak}.   
Let \boldmath$S$\unboldmath\, be the three-dimensional \textsc{Keccak} state that holds the 1600 state bits.
In the first step, the sum of every column is generated by \textsc{XOR}-ing the corresponding bits.
This generates the plane \boldmath$C$\unboldmath, accounting for 320 bits.
\begin{algorithm}[h]
    \caption{Theta Layer \cite{fipsnist}}
    \begin{flushleft}
    \hspace*{\algorithmicindent} \textbf{Input:} \textsc{Keccak} State Array \boldmath$S$\unboldmath \\
     \hspace*{\algorithmicindent} \textbf{Output:} \textsc{Keccak} State Array \boldmath$\theta$\unboldmath
    \end{flushleft}
    \begin{algorithmic}[1]
    \For{$x \gets 0$ to $4$, $z \gets 0$ to $63$}
        \State $C[x,z] \gets S[x,0,z] \oplus S[x,1,z] \oplus S[x,2,z] \oplus S[x,3,z] \oplus S[x,4,z]$
    \EndFor
    \For{$x \gets 0$ to $4$, $z \gets 0$ to $63$} 
        \State $D[x,z] \gets C[(x-1) \text{ mod 5}, z] \oplus C[(x+1) \text{ mod }5, (z-1) \text{ mod }63] $
    \EndFor
    \For{$x \gets 0$ to $4$, $y \gets 0$ to $4$, $z \gets 0$ to $63$}
        \State $\theta[x,y,z] \gets S[x,y,z] \oplus D[x,z]$
    \EndFor
    \State {Return \boldmath$\theta$\unboldmath}
    \end{algorithmic}
    \label{alg:theta}
\end{algorithm}

\begin{figure}[t]
\centerline{\includegraphics[width=.45\columnwidth]{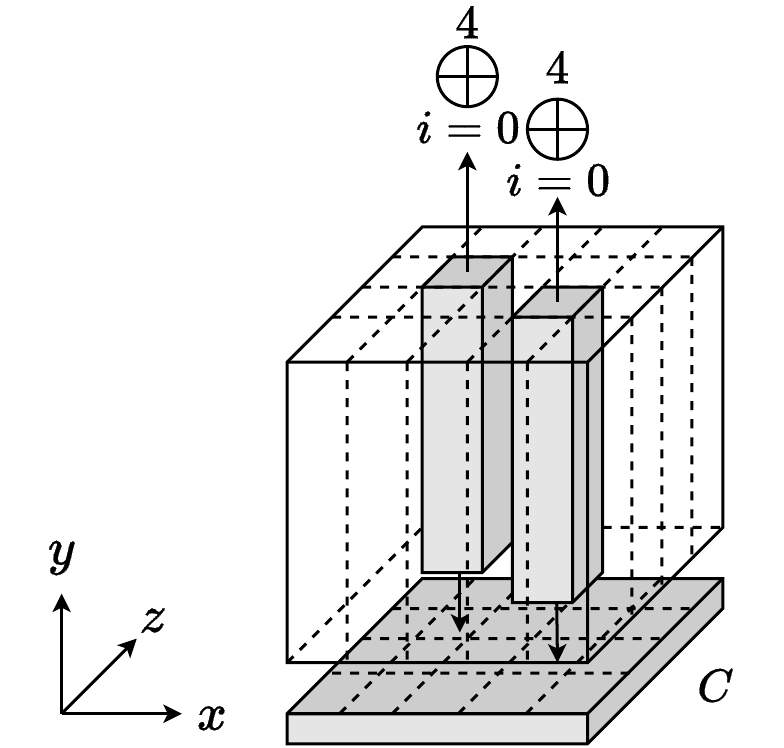}}
\caption{\textit{c-plane} Column Sum Generation}
\label{fig:c_plane_parity_check_only}
\end{figure}

After calculating the plane \boldmath$C$\unboldmath, the plane \boldmath$D$\unboldmath\, is generated by adding two diagonal columns nearby.
The output of the theta layer \boldmath$\theta$\unboldmath\, is calculated by adding a bit of the \boldmath$D$\unboldmath\, plane to all bits in the corresponding column of the \textsc{Keccak} state \boldmath$S$\unboldmath.
With our approach, we take advantage of established column sum generation performed in the theta layer and realize a FD based on parity checks. 
This solution exhibits low hardware and latency overhead.
In particular, we exploit the result of the \boldmath$C$\unboldmath\, plane computation, which is denoted by the \textit{c-plane} and shown in Fig.~\ref{fig:c_plane_parity_check_only}.
In detail, the \textit{c-plane}-based FD mechanism is described in Algorithm~\ref{alg:theta_protection_c_plane}:

\begin{algorithm}[h]
    \caption{\textit{c-plane}-based fault detection}\label{alg:theta_protection_c_plane}
    \begin{flushleft}
    \hspace*{\algorithmicindent} \textbf{Input:} \textit{c-plane} \boldmath$C$\unboldmath\, from the \textsc{Keccak} theta layer, \textsc{Keccak} state update \boldmath$S'$\unboldmath \\
    \hspace*{\algorithmicindent} \textbf{Output:} Fault detection signal $error$
    \end{flushleft}

\begin{algorithmic}[1]
\State $error \gets 0$

\LeftComment{Calculate the column sum bits \boldmath$C'$\unboldmath\, for comparison in the next \textsc{Keccak} round inside the protection unit}

\For{$x \gets 0$ to $4$, $z \gets 0$ to $63$}
    \State $C'[x,z] \gets S'[x,0,z] \oplus S'[x,1,z] \oplus S'[x,2,z] \oplus  S'[x,3,z] \oplus S'[x,4,z]$
\EndFor

\State \textbf{wait for} next clock cycle

\LeftComment{Compare the \textit{c-plane} \boldmath$C$\unboldmath\, from the \textsc{Keccak} theta layer and the stored \textit{c-plane} \boldmath$C'$\unboldmath\, for parity-checking}

\For{$x \gets 0$ to $4$, $z \gets 0$ to $63$}
    \State $error \gets error \lor (C[x,z]$ $\oplus$ $C'[x,z]$) \label{alg:theta_protection_cplane:line:error}
\EndFor
\State {Return $error$}

\end{algorithmic}
\end{algorithm}

When a \textsc{Keccak} permutation round is performed, the output signal of the \textsc{Keccak} round function \boldmath$S'$\unboldmath\, updates the \textsc{Keccak} state register.
The state-update signal \boldmath$S'$\unboldmath\, is routed into the FD unit to calculate the sum of all 320 columns and storing them in the register of \boldmath$C'$\unboldmath\, for comparison in the next \textsc{Keccak} permutation round.
While performing the next permutation round, the column sums \boldmath$C$\unboldmath\, are generated in the \textsc{Keccak} theta layer as described in Algorithm~\ref{alg:theta}.
Every bit of \boldmath$C$\unboldmath\, is compared with the corresponding pre-computed bit stored in the register of \boldmath$C'$\unboldmath. 
This approach provides a parity check for each column of the stored \textsc{Keccak} state \boldmath$S$\unboldmath.
If at least one of the bits in \boldmath$C$\unboldmath\, differs from the stored bits of \boldmath$C'$\unboldmath, the \textit{error} signal is raised to determine a detected fault.

\subsubsection{\textit{f-slice} Lane Sum Computation}
\label{subsec:f_slice_parity_check}

\begin{figure}[b]
    \centering
    \includegraphics[width=.5\columnwidth]{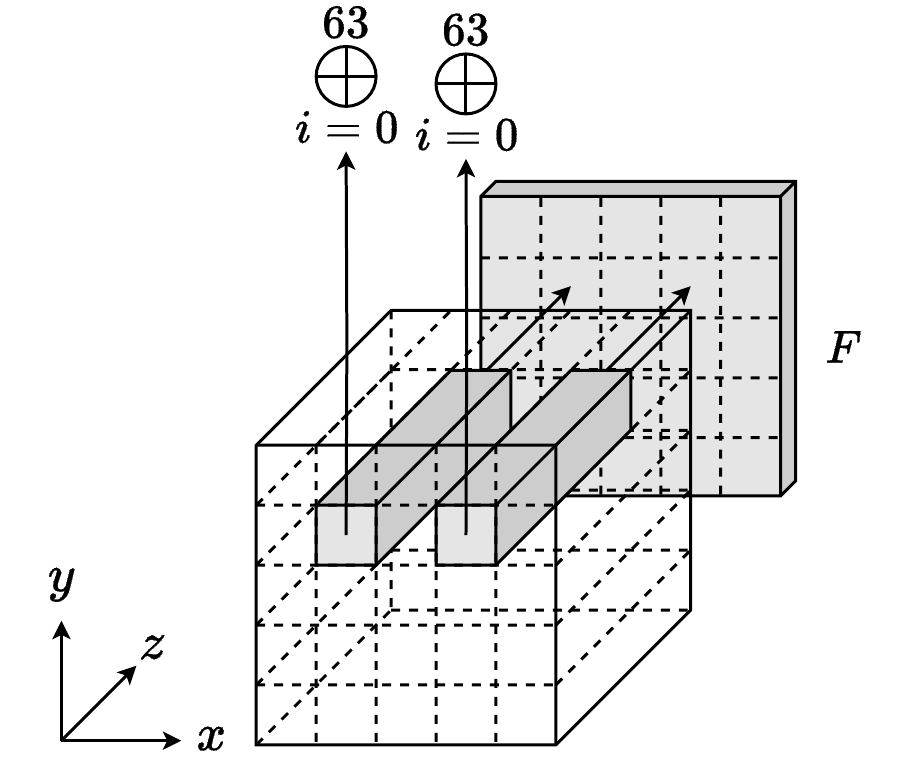}
    \caption{\textit{f-slice} Lane Sum Generation}
    \label{fig:f_slice_sum_generation}
\end{figure}

We extend the existing column sum calculation \textit{c-plane} present in the theta layer with a lane sum generation called \textit{f-slice} as shown in Fig.~\ref{fig:f_slice_sum_generation}. 

Algorithm~\ref{alg:theta_extension_fslice} describes \textit{f-slice} in details.  
Here, every lane's sum of the \textsc{Keccak} state is calculated, and the results generate a slice for comparison, accounting for an additional 25 bits.
\begin{algorithm}[h]
    \caption{\textit{f-slice} Lane Sum Computation}\label{alg:theta_extension_fslice}
    \begin{flushleft}
    \hspace*{\algorithmicindent} \textbf{Input:} \textsc{Keccak} State Array \boldmath$S$\unboldmath \\
     \hspace*{\algorithmicindent} \textbf{Output:} Lane Sum Slice \boldmath$F$\unboldmath
     \end{flushleft}
    \begin{algorithmic}[1]
        \State \boldmath$F$\unboldmath$ \gets 0$
        \For{$x \gets 0$ to $4$, $y \gets 0$ to $4$, $z \gets 0$ to $63$}
            \State $F[x,y] \gets F[x,y]$ $\oplus$ $S[x,y,z]$ 
        \EndFor
    \State {Return \boldmath$F$\unboldmath}
    \end{algorithmic}
\end{algorithm}

By adding the \textit{f-slice} to the theta layer and FD module, a parity-check similar to the \textit{c-plane}-based parity check is performed.
Combining the \textit{c-plane}- and \textit{f-slice}-based parity check enables a two-dimensional cross-parity check. 
The so-called \textit{z-sheet}-based parity-check provides a \textit{sheet}-based protection mechanism described as follows.

\subsubsection{\textit{z-sheet}-based Parity Check}
\label{subsubsec:z_sheet_parity_check}

\begin{figure}[tb]
\centerline{\includegraphics[width=.6\columnwidth]{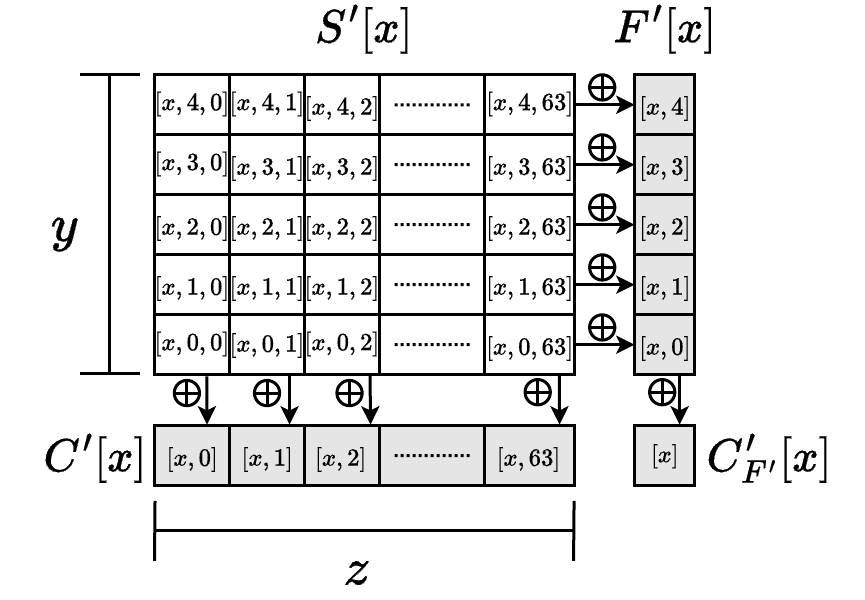}}
\caption{\textit{z-sheet} column and lane sum calculations carried out in the FD module}
\label{fig:two_dimensional_parity_check_sheet}
\end{figure}

The \textit{z-sheet}-based cross-parity check is described in Algorithm~\ref{alg:theta_protection_z_sheet}.
Similar to the \textit{c-plane}-only approach, the state update signal \boldmath$S'$\unboldmath\,serves as an input for the FD module.
In addition to the generation of the plane \boldmath$C'$\unboldmath, the slice \boldmath$F'$\unboldmath\,is calculated and the resulting bits are stored inside the FD module to be compared in the next \textsc{Keccak} round with the extended slice \boldmath$F$\unboldmath\, of the theta layer.
To protect the register of \boldmath$F'$\unboldmath, a parity check is added and performed as shown in Fig.~\ref{fig:two_dimensional_parity_check_sheet}.
During the calculation of \boldmath$F'$\unboldmath, the sum \boldmath$C'_{F'}$\unboldmath\, of every \boldmath$F'$\unboldmath s column is generated, accounting for another 5 sum bits.
Similar to the parity check of the \textsc{Keccak} state \boldmath$S$\unboldmath, the column-wise sum \boldmath$C_{F'}$\unboldmath\, of the \boldmath$F'$\unboldmath\, registers are calculated and compared to the bits previously stored in the registers of \boldmath$C'_{F'}$\unboldmath.

\begin{algorithm}[h]
    \caption{\textit{z-sheet}-based Fault Detection}\label{alg:theta_protection_z_sheet}
    \begin{flushleft}
    \hspace*{\algorithmicindent} \textbf{Input:} \textit{c-plane} \boldmath$C$\unboldmath\, \& \textit{f-slice} \boldmath$F$\unboldmath\, from the \textsc{Keccak} theta layer, \textsc{Keccak} state update \boldmath$S'$\unboldmath \\
    \hspace*{\algorithmicindent} \textbf{Output:} Fault detection signal $error$
    \end{flushleft}

\begin{algorithmic}[1]
\State \boldmath$F'$\unboldmath$ \gets 0$,  \boldmath$C'_{F'}$\unboldmath$ \gets 0$,  \boldmath$C_{F'}$\unboldmath$ \gets 0$, $error \gets 0$

\LeftComment{Calculate the column sum bits \boldmath$C'$\unboldmath\, for comparison in the next \textsc{Keccak} round}
\For{$x \gets 0$ to $4$, $z \gets 0$ to $63$} \label{alg:theta_protection_cplane:thetac1}
    \State $C'[x,z] \gets S'[x,0,z] \oplus S'[x,1,z] \oplus S'[x,2,z] \oplus  S'[x,3,z] \oplus S'[x,4,z]$ 
\EndFor
\LeftComment{Calculate the lane sum bits \boldmath$F'$\unboldmath\, for comparison in the next \textsc{Keccak} round}
    \For{$x \gets 0$ to $4$, $y \gets 0$ to $4$, $z \gets 0$ to $63$}
        \State $F'[x,y] \gets F'[x,y] \oplus S'[x,y,z]$ 
    \EndFor
\LeftComment{Calculate the sum of a \boldmath$F'$\unboldmath\, column for comparison in the next \textsc{Keccak} round}
    \For{$x \gets 0$ to $4$, $y \gets 0$ to $4$}
        \State $C'_{F'}[x] \gets C'_{F'}[x] \oplus F'[x,y]$ 
    \EndFor
    
\State \textbf{wait for} next clock cycle

\LeftComment{Calculate the sum of a \boldmath$F'$\unboldmath\, column for comparison}
    \For{$x \gets 0$ to $4$, $y \gets 0$ to $4$}
        \State $C_{F'}[x] \gets C_{F'}[x] \oplus F'[x,y]$ 
    \EndFor

\LeftComment{Compare the \textit{c-plane}s \boldmath$C$\unboldmath/\boldmath$C'$\unboldmath, \textit{f-slice}s \boldmath$F$\unboldmath/\boldmath$F'$\unboldmath\, and  \boldmath$F'$\unboldmath\, column sums \boldmath$C_{F'}$\unboldmath/\boldmath$C'_{F'}$\unboldmath\, for parity-checking}

\For{$x \gets 0$ to $4$, $y \gets 0$ to $4$, $z \gets 0$ to $63$}
    \State $error \gets error \lor (C[x,z] \oplus C'[x,z]) \lor (F[x,y] \oplus F'[x,y]) \lor (C_{F'}[x] \oplus C'_{F'}[x])$
\EndFor
\State {Return $error$}

\end{algorithmic}
\end{algorithm}

The \textit{z-sheet}-based parity check realizes a multidimensional parity check based on the \textit{sheets} of the \textsc{Keccak} state.
It should be noted that multidimensional cross-parity checks are an established technique for designing forward error correction (FEC) codes \cite{rubinoff1961, dudacek2016, vertat2019, jasim2019} for data transmission over unreliable data channels. 
Here, the transmitted data is represented in a multidimensional space to perform FEC over the state space.
However, a multidimensional cross-parity check for FEC is a costly technique from the hardware perspective.
The proposed two-dimensional cross-parity check approach provides a lightweight FD approach detecting up to three faults (bit flips) in the register of the \textsc{Keccak} state \boldmath$S$\unboldmath.

\subsection{Integration of the FD Mechanism into the Hash-Engine}
In the following, we show the integration of the proposed FD mechanism into the unified hash engine covering two implementation scenarios: round-based, and unrolled \textsc{Keccak} implementations.

\subsubsection{Round-based \textsc{Keccak} Implementation}
\label{subsec:integration_fault_detection_into_hash_engine}

\begin{figure}[b]
\centerline{\includegraphics[trim=1cm 0 0.8cm 0.1cm, clip, width=.8\columnwidth]{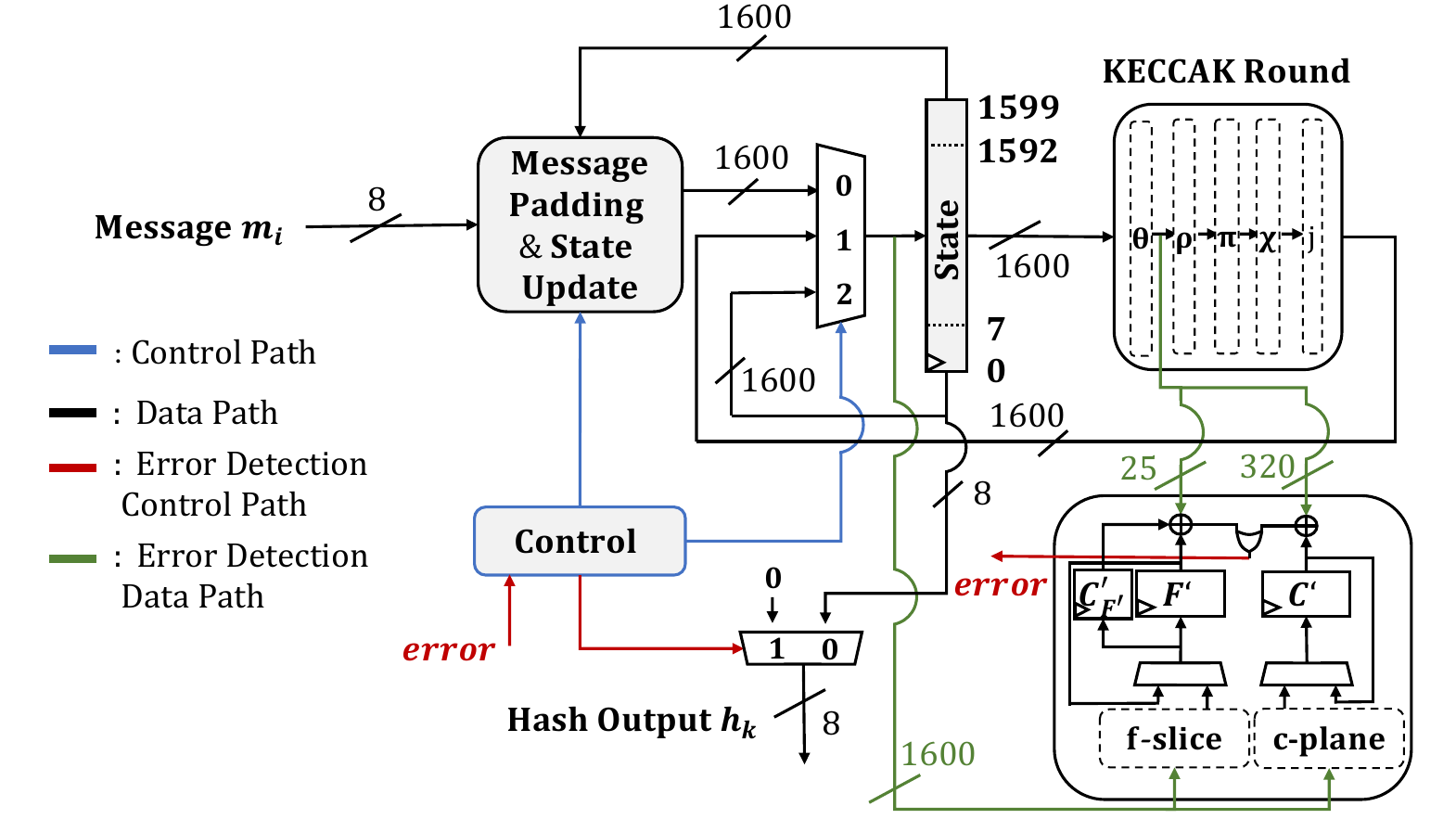}}
\caption{\textsc{Sha-3}/\textsc{Shake} FD Mechanism Design}
\label{fig:sha3_shake_protected_engine}
\end{figure}

We integrate the proposed FD mechanism stated in Section~\ref{sec:lightweight_fault_detection_mechanism} into the unified hash engine. 
The FD mechanism is implemented as a redundant module as depicted in Fig.~\ref{fig:sha3_shake_protected_engine}. 

The state-update signal \boldmath$S'$\unboldmath\, is forwarded as an input to the FD module.
Here, the \textit{c-plane} column and \textit{f-slice} lane sum calculation is performed and stored in the registers of \boldmath$C'$\unboldmath\, and \boldmath$F'$\unboldmath, respectively.
During the next computation cycle, the column sum \boldmath$C$\unboldmath\, and lane sum \boldmath$F$\unboldmath\ is calculated in the theta layer of the \textsc{Keccak} and forwarded to the FD module for comparison with the stored values in \boldmath$C'$\unboldmath\, and \boldmath$F'$\unboldmath.
In the case the \textit{z-sheet}-based cross-parity check is performed the column sum of \boldmath$F'$\unboldmath\, \textit{f-slice} is generated likewise. 
It is stored in the register of \boldmath$C'_{F'}$\unboldmath\, and compared by the generated column sum of \boldmath$F'$\unboldmath\, during the next computation cycle.
If any comparison fails, the column-wise \textit{c-plane}- or two-dimensional \textit{z-sheet}-based parity check mechanism detects a fault and the signal $error$ is raised.
As a result, the hash engine output is masked, preventing the leakage of the faulty digest.

It should be noted that both the \textit{c-plane}- and \textit{z-sheet}-based parity check focus on protecting the \textsc{Keccak} state registers. 
This work does not include any further protection of the unified hash engine's control unit and data interface.

\subsubsection{Unrolled \textsc{Keccak} Implementation}
\label{subsec:application_fault_detection_unrolled_keccak}

\begin{figure}[b]
\centerline{\includegraphics[trim=1cm 0 0.2cm 0, clip,width=.8\columnwidth]{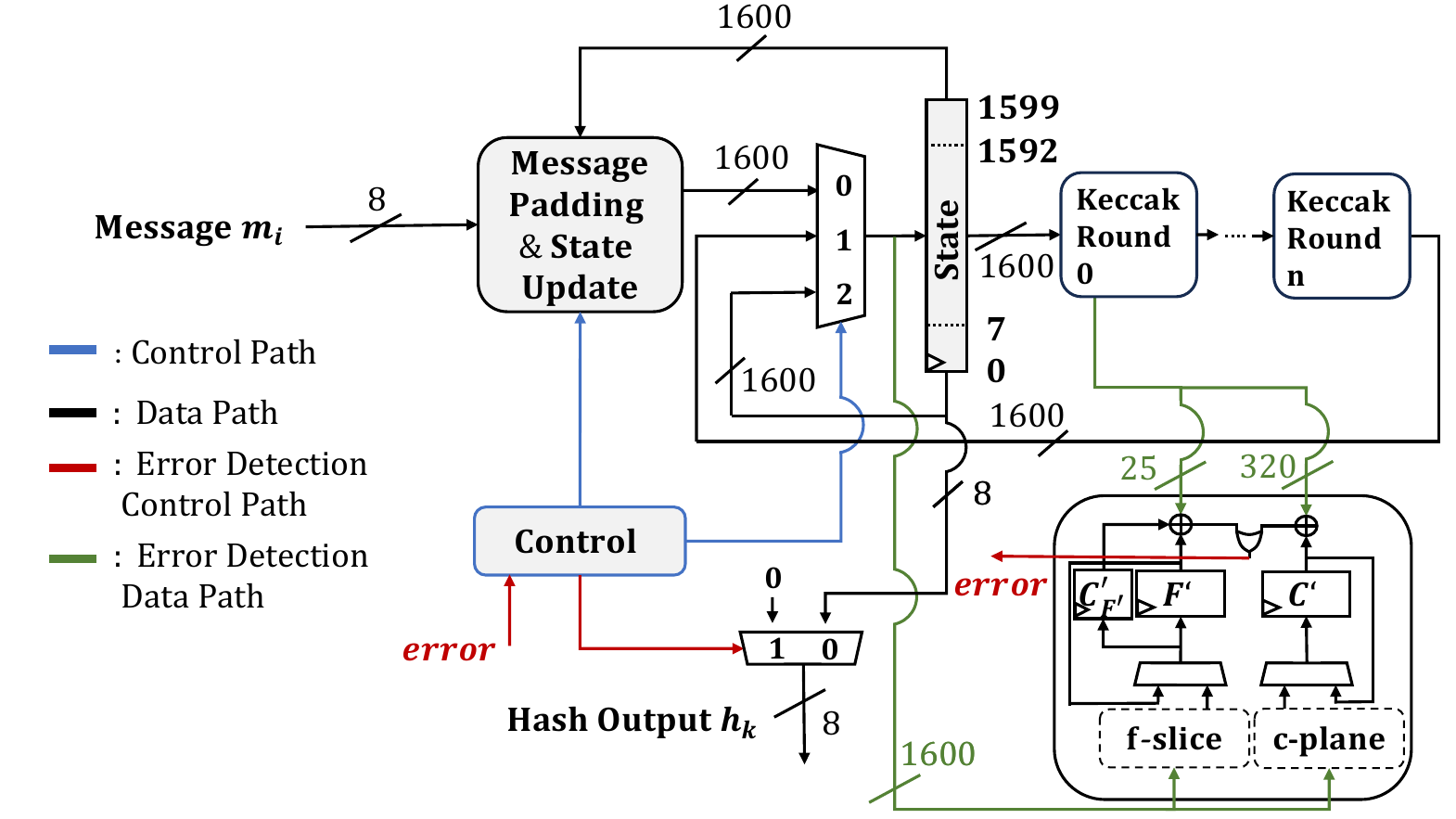}}
\caption{Unrolled \textsc{Sha-3}/\textsc{Shake}  Protected Hash Engine.}
\label{fig:sha3_shake_protected_engine_unrolled}
\end{figure}

As our FD mechanism relies on the \textsc{Keccak} state-update signal \boldmath$S'$\unboldmath\, to check the integrity of the state \boldmath$S$\unboldmath 's register, it applies to other realizations of the \textsc{Keccak}-$f$ permutation as long as the state is updated accordingly.
We show the flexibility of our approach not only by a round-based but also by different unrolled implementations.
For this, we modified our unified hash engine and realized unrolled implementations of the \textsc{Keccak}-$f$ permutation function as shown in Fig.~\ref{fig:sha3_shake_protected_engine_unrolled}.

Similar to the integration of the FD module into the round-based implementation of our unified hash engine, the state-update signal \boldmath$S'$\unboldmath\ serves as an input for the FD module, and the plane \boldmath$C'$\unboldmath\ is calculated.
In case of the \textit{z-sheet}-based cross-parity check, the slice \boldmath$F'$\unboldmath, and \boldmath$F'$\unboldmath s column sum \boldmath$C'_{F'}$\unboldmath\, is computed.
Accordingly, after the state register of \boldmath$S$\unboldmath\, is updated, the plane \boldmath$C$\unboldmath\, and slice \boldmath$F$\unboldmath\, from the theta layer of the first unrolled round is routed into the FD module for comparison.
Similar to the round-based implementation, the comparison of \boldmath$C_{F'}$\unboldmath\, and \boldmath$C'_{F'}$\unboldmath\, for the \textit{z-sheet}-based parity check is performed.

Based on the use case, our approach can be adapted from small round-based implementations to more sophisticated realizations with an unrolled design.
Furthermore, our FD mechanism applies to unrolled and pipelined implementations deployed in high-performance applications.
In the latter case, the FD module is applied to every register holding the \textsc{Keccak} state \boldmath$S$\unboldmath\, and reports detected faults.
The proposed unified hash engine is intended to be deployed in resource-constrained devices.
Pipelined fault-resilient \textsc{Keccak} implementations would have a huge impact on the required hardware, thus we do not consider such an implementation in this work.

\subsection{Integrating the Unified Hash Engine into a RISC-V SoC}
\label{subsec:integration_riscv_soc}
\begin{figure}[t]
\centerline{\includegraphics[width=.5\columnwidth]{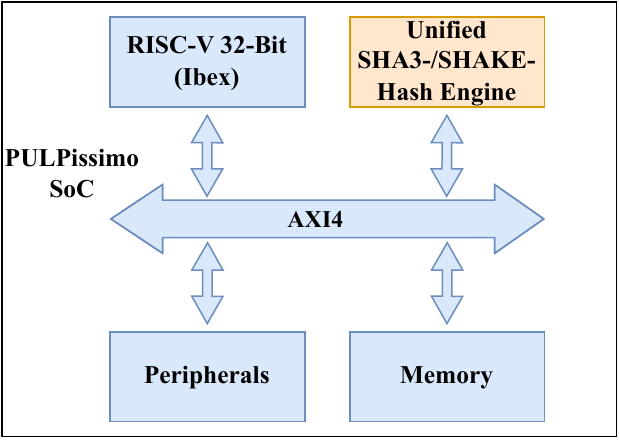}}
\caption{Block Diagram of 32-bit RISC-V Microcontroller with the Proposed Unified Hash Engine}
\label{fig:SoCPQC}
\end{figure}

The unified hash engine's design provides a byte-wise input/output interface that can be connected to a bus interface, like AXI4 or Tilelink. 
Therefore, the engine can be integrated into any System-on-Chip (SoC) as a memory-mapped IP. 
The hash engine is to be integrated into a low-power 32-bit RISC-V microcontroller, as shown in Fig.~\ref{fig:SoCPQC}, to be comparable with the SoA hardware used in PQC.    

\section{Implementation Results}
\label{sec:results}
This section shows the implementation results of the \textsc{Sha3}/\textsc{Shake} unified hash engine and the integration of the proposed FD mechanism and compares its security level to SoA fault injection countermeasures for \textsc{Keccak}.
Then, we compare our implementation results for ASIC and FPGA technology to SoA. 
We also showcase the performance and the integration results into the PULPissimo SoC of our design.

\subsection{The Unified Hash Engine Implementation}
\label{subsec:implementation_results_round_based_engine}
The proposed \textsc{Sha3}/\textsc{Shake} engine is written in VHDL and synthesized for 45\,nm FreePDK45 \cite{freepdk45} standard library using Synopsys Design Compiler U-2022.12 \cite{synopysdc}.
We attached the proposed engine via AXI4 to the PULPissimo SoC \cite{pulpissimo2018} as a memory-mapped device to demonstrate integration into a lightweight SoC.
We also implemented the design on the Xilinx ZCU102 evaluation board containing a Virtex-7 FPGA using Xilinx Vivado 2022.1 \cite{vivado}.

\subsubsection{Round-based Implementation}
In the following, we discuss the implementation results of the proposed \textsc{Sha3}/\textsc{Shake} engine in ASIC and FPGA technologies.
It covers three cases: the unprotected, the \textit{c-plane}, and the \textit{z-sheet} protected engine.  

\paragraph{Round-based FPGA Implementation Results}
\label{subsubsec:comparison_fpga_results}
Table~\ref{tab:fpga_comparision_roy2020high} shows the FPGA implementation results compared to the SoA unified \textsc{Sha3}/\textsc{Shake} engines of \cite{roy2020high} and \cite{xi2023low} based on FPGA primitives, namely Look-up-Tables (LUT), Flip-Flops (FF), and DSP; the latter denotes a cluster including LUTs and FFs.

\begin{table}[h]
\centering
\caption{FPGA Implementation Comparison of Unified \textsc{SHA-3}/\textsc{Shake} Engines}
\label{tab:fpga_comparision_roy2020high}
\begin{tabular}{c|ccc|c}
\multirow{2}{*}{\textbf{Paper}} & \multicolumn{3}{c|}{\textbf{Area}} & \multicolumn{1}{c}{\multirow{2}{*}{\textbf{FPGA}}} \\
                                & LUT       & FF        & DSP     & \multicolumn{1}{c}{}                               \\ \hline
\textbf{\cite{roy2020high}}     & 5113      & 3068      & -          & Ultrascale+                                        \\
\textbf{\cite{xi2023low}}       & 7376      & 3059      & 4       & Kintex-7         \\
\textbf{[This work]}            & 3566      & 1645      & -        & Virtex-7                               
\end{tabular}
\end{table}

Although our design of a unified \textsc{Sha3}/\textsc{Shake} engine covers all standard hash modes: \textsc{Sha-3-224}, \textsc{Sha-3-256}, \textsc{Sha-3-384}, \textsc{Sha-3-512}, \textsc{Shake-128}, and \textsc{Shake-256} while the design of \cite{roy2020high} and \cite{xi2023low} only realizes three hash modes, our design requires less area compared to \cite{roy2020high}, due to our proposal for byte-wise in-place state partitioning. 
This shows the flexibility and lightweightness of our unified \textsc{Sha3}/\textsc{Shake} engine. 

For general FPGA comparison with the SoA \textsc{Keccak} implementations,  Table~\ref{tab:fpga_results_comparison} summarizes the FPGA implementation results. Our \textit{c-plane}-based resilient hash engine requires $2.07\times$ fewer slices than the SoA solutions detecting up to one bit flip \cite{kahri2017, mestiri2023}.
Compared to SoA approaches, detecting up to three-bit flips  \cite{gavrilan_2024}, our \textit{z-sheet}-based FD mechanism shows an improvement factor of 2.4 regarding LUT overhead (89\% vs. 214\%) and a 5$\times$ FF overhead improvement (20\% vs. 100\%). 
For the protected design, this results in overall $1.8\times$ fewer LUTs and $1.6\times$ fewer FFs.
This underlines our FD mechanism as a lightweight, resilient solution that is competitive with the state of the art.
\begin{table}[b]
\caption{Round-based FPGA Implementation Results Comparison}
\label{tab:fpga_results_comparison}
\begin{tabular}{cl|cccc|cccc|c|c}
\toprule
\multicolumn{2}{c|}{\multirow{2}{*}{\textbf{Paper}}}                                          
    & \multicolumn{4}{c|}{\textbf{without protection}}  
    & \multicolumn{4}{c|}{\textbf{with protection}}                                      
    & \multicolumn{1}{c|}{\multirow{2}{*}{\textbf{FPGA}}}
    & \multirow{3}{*}{%
        \textbf{\begin{tabular}[c]{@{}c@{}} Fault\\ Detectability\\ (bit flips) \end{tabular}}%
    } \\
\multicolumn{2}{c|}{}
& LUT  & FF   & Slices & 
\begin{tabular}[c]{@{}c@{}}Freq.\\ (MHz)\end{tabular} 
& LUT  & FF    & Slices & 
\begin{tabular}[c]{@{}c@{}}Freq.\\ (MHz)\end{tabular} 
& \multicolumn{1}{c|}{}\\ \midrule
\multicolumn{2}{c|}{\textbf{\cite{gavrilan_2024}}}                                                                    & 3953 & 1621 & 1036   & 222.2                                                     & 12408 & 3260  & 3269   & 111.10                                                    & Artix-7 & $\leq$ 3                                             \\
\multicolumn{2}{c|}{\textbf{\begin{tabular}[c]{@{}c@{}}\cite{torresalvardo2022}\\ ArchHC\end{tabular}}}     & 2339 & 2361 & -      & 228.25                                                    & 28703 & 18192 & -      & 45.89                                                     & Virtex-7 & $-$                                          \\
\multicolumn{2}{c|}{\textbf{\begin{tabular}[c]{@{}c@{}}\cite{torresalvardo2022}\\ ArchTMR\_HC\end{tabular}}} & 2339 & 2361 & -      & 228.25                                                    & 27226 & 26197 & -      & 63.58                                                     & Virtex-7 & $-$                      \\
\multicolumn{2}{c|}{\textbf{\cite{mestiri2023}}}                                                                      & -    & -    & 1370   & 258.60                                                    & -     & -     & 1680   & 387.00                                                    & Virtex-5 & $\leq$ 1                                          \\
\multicolumn{2}{c|}{\textbf{\cite{kahri2017}}}                                                                        & -    & -    & 1356   & 296.50                                                    & -     & -     & 2260   & 291.30                                                    & Virtex-5 & $\leq$ 1                                          \\
\multicolumn{2}{c|}{\textbf{\cite{mestiri2021}}}                                                                      & -    & -    & 1350   & 252.40                                                    & -     & -     & 1601   & 365.2                                                     & Virtex-5 & $<$ 1                                           \\
\multicolumn{2}{c|}{\textbf{\begin{tabular}[c]{@{}c@{}}[This work]\\ c-plane\end{tabular}}}                                & 3566 & 1645 & 563    & 20.00                                                         & 4868  & 1966  & 809   & 20.00                                                         & Virtex-7 &  $\leq$ 1                                         \\
\multicolumn{2}{c|}{\textbf{\begin{tabular}[c]{@{}c@{}}[This work]\\ z-sheet\end{tabular}}}                     & 3566 & 1645 & 563    & 20.00                                                         & 6738  & 1996  & 1067   & 20.00                                                         & Virtex-7 & $\leq$ 3                                         
\end{tabular}
\end{table}

\paragraph{Round-based ASIC Implementation Results} 
\label{subsubsec:comparison_asic_results}

\setlength{\tabcolsep}{2pt}
\begin{table}[h]
\caption{Round-based ASIC Implementation Results Comparison}
\label{tab:asic_results_comparison}
\begin{tabular}{cl|cc|cc|c|c}
\toprule
\multicolumn{2}{c|}{\multirow{2}{*}{\textbf{Paper}}}                                                   & \multicolumn{2}{c|}{\textbf{without protection}}                                                                            & \multicolumn{2}{c|}{\textbf{with protection}}                                                                               & \multicolumn{1}{c|}{\multirow{2}{*}{\textbf{PDK}}} & 
\multirow{2}{*}{\textbf{\begin{tabular}[c]{@{}c@{}} Fault\\ Detectability \\ (bit flips) \end{tabular}}} \\
\multicolumn{2}{c|}{}                                                                                  & \begin{tabular}[c]{@{}c@{}}Area\\ (kGE)\end{tabular} & \begin{tabular}[c]{@{}c@{}}Freq.\\ (MHz)\end{tabular} & \begin{tabular}[c]{@{}c@{}}Area\\ (kGE)\end{tabular} & \begin{tabular}[c]{@{}c@{}}Freq.\\ (MHz)\end{tabular} & \multicolumn{1}{c|}{}                              \\ \midrule
\multicolumn{2}{c|}{\textbf{\cite{gavrilan_2024}}}                                                           & 57.10                                               & 1.316,00                                                    & 177.70                                                & 719.40                                                    & FreePDK45  & $\leq$ 3                                   \\
\multicolumn{2}{c|}{\textbf{\cite{luo2016}}}                                                           & 52.14                                                & 256.94                                                    &  83.48                                               & 228.26                                                    & FreePDK45 & $\leq$ 1                                  \\
\multicolumn{2}{c|}{\textbf{\cite{bayatsarmadi2014}}}                                                  & 45.40                                                        & 676.00                                                    & 49.10                                                        & 1192.00                                                   & 65nm TSMC &  $\leq$ 1                                        \\
\multicolumn{2}{c|}{\textbf{\begin{tabular}[c]{@{}c@{}}\cite{purnal2019}\\ KIT\end{tabular}}}          & -                                                               & -                                                         & 89.60                                                        & -                                                         & FreePDK45 &  $<$ 1                                   \\
\multicolumn{2}{c|}{\textbf{\begin{tabular}[c]{@{}c@{}}\cite{purnal2019}\\ FUR\end{tabular}}}          & -                                                               & -                                                         & 177.20                                                       & -                                                         & FreePDK45 & $<$ 1                                   \\
\multicolumn{2}{c|}{\textbf{\begin{tabular}[c]{@{}c@{}}\cite{purnal2019}\\ FAST\end{tabular}}}         & -                                                               & -                                                         & 1317.80                                                      & -                                                         & FreePDK45 & $<$ 1                                  \\
\multicolumn{2}{c|}{\textbf{\begin{tabular}[c]{@{}c@{}}\cite{purnal2019}\\ BLAZE\end{tabular}}}        & -                                                               & -                                                         & 2494.60                                                      & -                                                         & FreePDK45 & $<$ 1                                   \\
\multicolumn{2}{c|}{\textbf{\begin{tabular}[c]{@{}c@{}}[This work]\\ c-plane\end{tabular}}}            & 25.65                                                        & 714.29                                                    & 33.17                                                           & 666.67                                                    & FreePDK45 & $\leq$ 1                                   \\
\multicolumn{2}{c|}{\textbf{\begin{tabular}[c]{@{}c@{}}[This work]\\ z-sheet\end{tabular}}} & 25.65                                                        & 714.29                                                    & 39.96                                                           & 588.24                                                    & FreePDK45 & $\leq$ 3                                  
\end{tabular}
\end{table}

We compare different SoA implementations by using the concept of \textit{gate equivalent} (GE) as shown in Table~\ref{tab:asic_results_comparison}. 
The results show that our protected design requires less area than the SoA fault detection approaches for one bit-flip.
The \textit{c-plane}-protected hash engine exhibits $1.48\times$ smaller area requirements than the SoA solutions detecting up to one bit flip \cite{luo2016, bayatsarmadi2014}.
Compared to the SoA solution detecting up to three-bit flips \cite{gavrilan_2024}, our \textit{z-sheet}-based FD mechanism itself shows an improvement in area overhead by $3.7\times$ ($56\,\%$ vs. $211\,\%$).
This improvement results in an overall $4.5\times$ smaller hash-engine design while achieving the same fault detectability.

\subsubsection{Unrolled Implementation}
The following discusses the unrolled implementation results of the proposed \textsc{Sha3}/\textsc{Shake} engine in ASIC and FPGA technologies.
It covers several unrolled implementation cases, ranging from two to 24 rounds.

\paragraph{Unrolled FPGA Implementation Results}
Table~\ref{tab:FPGA_results_unrolled} shows the FPGA results of the unrolled implementation of the unified engine with/without the proposed FD mechanism. 
\textit{Z-sheet} protection causes area overhead from 50.7\% for two unrolled rounds to just 6.56\% for a fully unrolled engine.
\begin{table}[h]
\caption{Unrolled FPGA Implementation Results}
\label{tab:FPGA_results_unrolled}
\begin{tabular}{c|ccc|ccc|cccc}
\hline
\multirow{2}{*}{\begin{tabular}[c]{@{}c@{}}\textbf{$\#$ Unrolled} \\ \textbf{Rounds}\end{tabular}} & \multicolumn{3}{c|}{\textbf{w/o protection}} & \multicolumn{3}{c|}{\textbf{w/ \textit{c-plane} protection}} & \multicolumn{4}{c}{\textbf{w/ \textit{z-sheet} protection}}                           \\
                                               & LUT           & FF           & Slices        & LUT                & FF                & Slices              & LUT   & FF   & Slices & \begin{tabular}[c]{@{}c@{}}Slices \\ Overhead\end{tabular} \\ \hline
2                                              & 5694          & 1645         & 869           & 8358               & 1966              & 1273                & 8862  & 1996 & 1310   & 50.7\,\%                                                      \\
4                                              & 9930          & 1645         & 1460          & 12140              & 1966              & 1869                & 14637 & 1996 & 2318   & 58.7\,\%                                                      \\
6                                              & 14153         & 1645         & 2074          & 21120              & 1966              & 3162                & 19477 & 1996 & 2946   & 42.02\,\%                                                     \\
8                                              & 18384         & 1645         & 2784          & 21853              & 1966              & 3223                & 20479 & 1996 & 3199   & 14.9\,\%                                                      \\
12                                             & 26836         & 1645         & 4452          & 28759              & 1966              & 4677                & 31194 & 1996 & 5083   & 14.1\,\%                                                      \\
24                                             & 52131         & 1645         & 8625          & 56764              & 1966              & 9333                & 56503 & 1996 & 9191   & 6.56\,\%                                                     
\end{tabular}
\end{table}

The FPGA implementation demonstrates that increasing the amount of unrolling leads to a significant decrease in \textit{z-sheet} protection area overhead. 

\paragraph{Unrolled ASIC Implementation Results} 
Table~\ref{tab:asic_results_unrolled} shows the impact of the proposed FD mechanism on unrolled implementations of the unified hash engine. 

\begin{table}[h]
\caption{ASIC results for unrolled implementations}
\label{tab:asic_results_unrolled}
\begin{tabular}{c|cc|cc|ccc}
\hline
\multirow{2}{*}{\begin{tabular}[c]{@{}c@{}}\textbf{$\#$ Unrolled} \\ \textbf{Rounds}\end{tabular}} & \multicolumn{2}{c|}{\textbf{w/o protection}}                                                                 & \multicolumn{2}{c|}{\textbf{w/ \textit{c-plane} protection}}                                                 & \multicolumn{3}{c}{\textbf{w/ \textit{z-sheet} protection}}                                                                                    \\
                                             & \begin{tabular}[c]{@{}c@{}}Area\\ (kGE)\end{tabular} & \begin{tabular}[c]{@{}c@{}}Freq,\\ (MHz)\end{tabular} & \begin{tabular}[c]{@{}c@{}}Area\\ (kGE)\end{tabular} & \begin{tabular}[c]{@{}c@{}}Freq.\\ (MHz)\end{tabular} & \multicolumn{2}{c}{\begin{tabular}[c]{@{}c@{}}Area \\ (kGE) (\% Overhead)\end{tabular}} & \begin{tabular}[c]{@{}c@{}}Freq.\\ (MHz)\end{tabular} \\ \hline
2                                            & 45.57                                                & 617.28                                                & 53.81                                                & 540.54                                                & 60,04                                     & 31.7\,\%                                   & 423,73                                                \\
4                                            & 64.06                                                & 403.23                                                & 73.57                                                & 367.65                                                & 75,90                                     & 18.4\,\%                                   & 280,11                                                \\
6                                            & 89.17                                                & 276.24                                                & 93.01                                                & 202.43                                                & 109,97                                    & 23.3\,\%                                   & 223,21                                                \\
8                                            & 121,24                                               & 201,21                                                & 130.40                                               & 193.05                                                & 135,92                                    & 12.1\,\%                                   & 176,37                                                \\
12                                           & 175.75                                               & 135.69                                                & 182.25                                               & 131.75                                                & 190,57                                    & 8.4\,\%                                    & 124,69                                                \\
24                                           & 314.76                                               & 72.25                                                 & 333.15                                               & 71.28                                                 & 339,06                                    & 7.7\,\%                                    & 68,45                                                
\end{tabular}
\end{table}
The proposed FD mechanism (\textit{z-sheet}) itself leads to an area overhead depending on the amount of unrolling, ranging from 31.7\% (two rounds) to 7.7\% (24 rounds).
With increasing numbers of unrolled rounds, the \textit{z-sheet} protection overhead significantly decreases.         
As a result, a fully unrolled engine (24 rounds) only shows 7.7\% area overhead for applied \textit{z-sheet} protection.

\subsection{Power \& Throughput Analysis}
\label{subsec:performance_efficiency_eval}

We evaluate the FD mechanism's impact on the round-based unified hash engine from power and throughput perspectives.
For this, we synthesized the unified hash engine and FD module for the 45\,nm FreePDK45 standard library \cite{freepdk45}.
Table~\ref{tab:power} shows our design's power consumption for three configurations: unprotected, protected with \textit{c-plane}, and protected with \textit{z-sheet}.
\begin{center}
    \begin{table}[ht]
\caption{Power for unprotected and protected designs based on 45\,nm FreePDK45 results}
\label{tab:power}
\centering
\begin{tabular}{cc}
\toprule
{\textbf{Design}} & \textbf{Power / mW} \\ \midrule
{\textbf{\textbf{Basic}}}         &         9.50                          \\ 
{\textbf{\textbf{w/ c-plane}}}         &         8.63                     \\ 
{\textbf{\textbf{w/ z-sheet}}}         &         9.05            \\ \bottomrule
\end{tabular}
\end{table}
\end{center}
The \textit{c-plane} protected design requires lower power than the unprotected design due to the decreasing clock frequency.
In the case of \textit{z-sheet} protection, the combinational logic and register increases, and more power is consumed compared to the \textit{c-plane}-based protection.
The throughput results presented in Fig.~\ref{fig:throughput} are calculated based on the hash modes.
\definecolor{bblue}{HTML}{4F81BD}
\definecolor{rred}{HTML}{C0504D}
\definecolor{ggreen}{HTML}{9BBB59}
\definecolor{ppurple}{HTML}{9F4C7C}

\begin{figure}[ht]
\begin{tikzpicture}
    \begin{axis}[
        width  = .95\columnwidth,
        height = .45\columnwidth,
        minor tick num=1,
        minor x tick style = transparent,
        major x tick style = transparent,
        ybar=2*\pgflinewidth,
        bar width=11pt,
        ymajorgrids = true,
        yminorgrids = true,
        ylabel = {Throughput / Mbps},
        symbolic x coords={{Sha3-224}, {Sha3-256}, {Sha3-384}, {Sha3-512}, {Shake128}, {Shake256}},
        xtick = data,
        x tick label style={font=\scshape,align=center},
        scaled y ticks = true,
        enlarge x limits=0.10,
        ymin=0,
        legend cell align=left,
        legend style={
                at={(1,1.05)},
                anchor=south east,
                column sep=1ex
        }
    ]
        \addplot[style={bblue,fill=bblue,mark=none}]
            coordinates {(Shake128, 4998.90) (Shake256, 4046.20) (Sha3-224, 4284.32) (Sha3-256, 4046.20) (Sha3-384, 3094.00) (Sha3-512, 2142.07)};

        \addplot[style={rred,fill=rred,mark=none}]
            coordinates {(Shake128, 4665.61) (Shake256, 3776.43) (Sha3-224,  3998.67) (Sha3-256, 3776.43) (Sha3-384, 2887.72) (Sha3-512, 1999.26)};

        \addplot[style={ggreen,fill=ggreen,mark=none}]
            coordinates {(Shake128, 4116.71) (Shake256, 3332.14) (Sha3-224, 3528.24) (Sha3-256, 3332.14) (Sha3-384, 2547.99) (Sha3-512, 1764.05)};

        \legend{Basic Design,w/ \textit{c-plane},w/ \textit{z-sheet}}
    \end{axis}
\end{tikzpicture}
\caption{Throughput of the hash-modes based on 45nm FreePDK45 results}
\label{fig:throughput}
\end{figure}
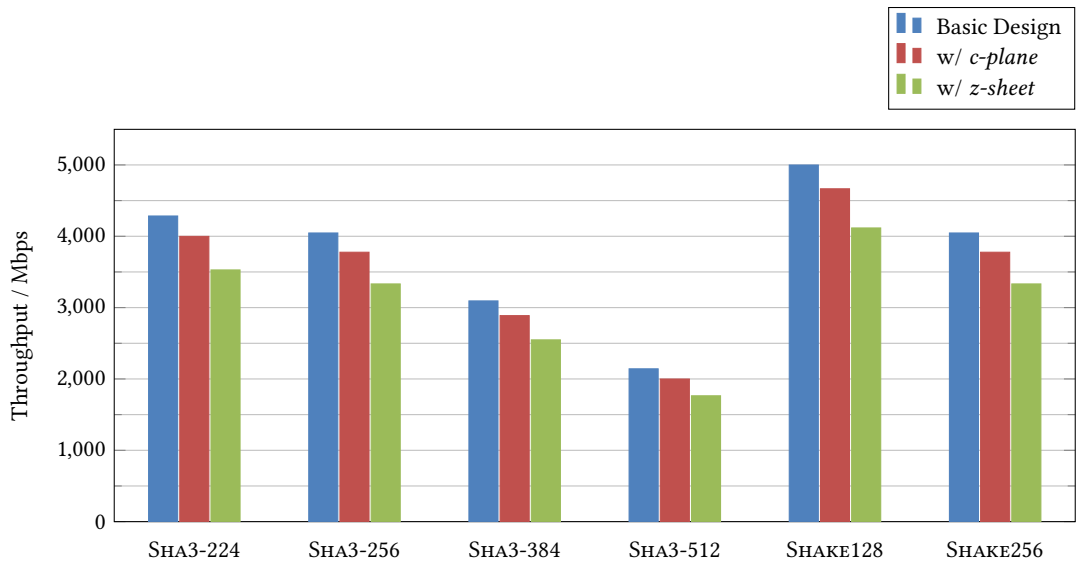
As expected from the lowered maximum clock frequency, the throughput for \textit{c-plane} and \textit{z-sheet} protected designs are lower than for the basic design. However, as outlined in Section~\ref{subsec:performance_efficiency_eval}, the throughput increases with hash modes using a higher rate size for operating on byte granularity and in-place updating of the (padded) messages.
\subsection{Integration into PULPissimo System-on-Chip}
\label{subsubsec:pulpissimo_integration_results}

For showcasing our design's applicability to a lightweight SoC running PQC applications
, we integrated our unified hash engine as a round-based implementation into the PULPissimo SoC \cite{pulpissimo2018}.
The engine is attached to the system bus as a memory-mapped device using AXI4 as illustrated in Fig.~\ref{fig:SoCPQC}.
For analysis, we deployed PULPissimo's FPGA toolflow \cite{pulpissimogithub}, which implements the design on a the Virtex-7 FPGA for a 20\,MHz clock frequency.
The results are presented in Table~\ref{tab:area_consumption_pulpissimo}.
Registers consume high area overhead compared to logic elements, hence, the amount of registers in our design is kept as low as possible, leading to a minimal register overhead of 3.9\,\% for the unprotected hash engine integrated in the PULPissimo SoC.
With applied protection mechanisms, the register overhead increases to 4.7\% for the \textit{z-sheet}-based protection.
Furthermore, an overhead of $<$13$\%$ for LUTs and $<$8$\%$ for slices is observed.
These results emphasize the low area overhead of the resulting fault-resilient unified hash engine, thus making it suitable for microcontroller-based and embedded IoT applications.

\begin{table}[t]
\caption{Hash Engine Area Overhead for PULPissimo SoC on Virtex-7 FPGA}
\label{tab:area_consumption_pulpissimo}
\centering
\begin{tabular}{@{}ll|c|c|c@{}}
\toprule
\multicolumn{2}{l|}{} & \multicolumn{3}{c}{\textbf{Protection Mechanism}} \\
\multicolumn{2}{l|}{\textbf{\textbf{Virtex-7} $@$ 20MHz}} & \textbf{none} & \textbf{c-plane} & \textbf{z-sheet}         \\ \midrule
               & LUTs (SoC)                                     &       52976               &     54046        &       55538       \\  
               & LUTs (Hash Engine)                             &       3566                &     4868         &       6738        \\
               & $\%$ (Hash Engine)                             &       6.73\,\%                &     9.00\,\%         &       12.13\,\%       \\ \midrule
               & Flip-flops (SoC)                               &       42317               &     42638        &       42668       \\
               & Flip-flops (Hash Engine)                       &       1645                &     1966         &       1996        \\
               & $\%$ (Hash Engine)                             &       3.89\,\%                &     4.61\,\%         &       4.68\,\%        \\ \midrule
               & Slices (SoC)                                   &       13247               &     13662        &       13849       \\
               & Slices (Hash Engine)                           &       563                 &     809          &       1067         \\
               & $\%$ (Hash Engine)                             &       4.25\,\%                &     5.92\,\%         &       7.70\,\%        \\ \bottomrule
\end{tabular}
\end{table}

\subsection{Fault Detectability Analysis and Comparison}
\label{subsec:comparison_existing_implementations}

Multiple fault countermeasures and protection schemes have been proposed in the literature. 
Depending on whether the fault is detected or corrected, these countermeasures are mainly classified into fault detection and correction schemes \cite{potestad2022hardware}.
These fault-tolerance techniques can be grouped into three main categories: Spatial redundancy, temporal redundancy, and information redundancy \cite{potestad2022hardware}.
Therefore, we compare the proposed FD mechanism to SoA implementations presented in Fig.~\ref{fig:sota_overview}. 

\usetikzlibrary{trees, arrows, positioning}
\tikzstyle{level 1}=[level distance=15mm, sibling distance=12mm]
\tikzstyle{level 2}=[level distance=15mm, sibling distance=5mm]
\begin{figure}[b]
\centering
\begin{tikzpicture}[child anchor=west, grow=east,->,>=angle 60]
\begin{scope}[yshift=0]
  \node [anchor=east] {Redundancy}
    child {
        node [anchor=west] {Information}
            child {node [anchor=west, xshift=0mm] {FPGA\cite[This work]{gavrilan_2024, torresalvardo2022}}}
            child {node [anchor=west, xshift=0mm] {ASIC\cite[This work]{gavrilan_2024, luo2016, purnal2019}}}
        }
    child {
        node [anchor=west] {Temporal}
            child {node [anchor=west, xshift=1.5mm] {FPGA\cite{mestiri2021, mestiri2023}}}
            child {node [anchor=west, xshift=1.5mm] {ASIC\cite{bayatsarmadi2014}}}
        }
    child {
        node [anchor=west] {Spatial}
            child {node [anchor=west, xshift=3mm] {FPGA\cite[This work]{kahri2017, mestiri2023, torresalvardo2022}}}
            child {node [anchor=west, xshift=3mm] {ASIC[This work]}}
        }
    ;
\end{scope}
\end{tikzpicture}
\caption{Overview of State of the Art implementations}
\label{fig:sota_overview}
\end{figure}
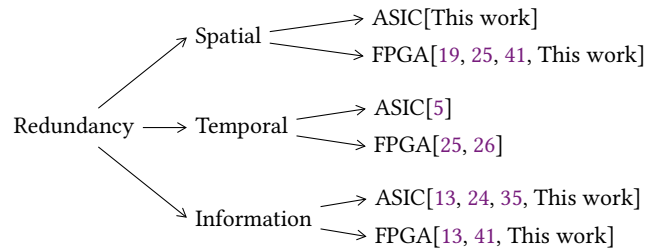

The comparison is performed as follows: 
\begin{itemize}
    \item \textbf{Spatial Redundancy}:
    This type is based on replicating logic/state and performing redundant computation at the hardware level.
    In \cite{kahri2017}, an FD mechanism for \textsc{Keccak} based on a double-redundant scrambling technique was proposed. 
    This mechanism is effective against single bit-flips with a fault coverage of 99.996\,\%.
    According to Table~\ref{tab:fpga_results_comparison}, the proposed mechanism in \cite{kahri2017} exhibits high overhead with even fault detectability compared to our work.  
    \item \textbf{Temporal Redundancy}: This type is based on repeating the operation twice on the same logic with the output being compared between both runs.
    Although less cost-intensive in terms of area, this type of redundancy decreases the performance by $50$\,$\%$ on average for performing each operation twice.
    In \cite{bayatsarmadi2014}, a recalculation with rotated operands was performed to provide a time-redundant FD scheme. 
    This mechanism is efficient only against a single bit-flip.
    Another work provides a time-redundancy-based scheme by splitting the \textsc{Keccak} round into two halves rounds through the pipeline register \cite{mestiri2021}.
    While the result of the first half-round is fed to the second half of the \textsc{Keccak} round, the input of the first half is re-executed and compared with the result stored in a pipeline register. In case of a mismatch, an error is raised.
    This approach is deployed to the second half of \textsc{Keccak} round to protect the complete design.
    The performed fault analysis shows a fault-detection probability of 99.458\,\% for a single bit-flip, thus not providing 100\,\% detection capability.
    Both temporal redundant mechanisms of \cite{bayatsarmadi2014} and \cite{mestiri2021} show a huge impact on the throughput compared to our work.
    \item \textbf{Information Redundancy}:
    This type is based on incorporating redundant information in the form of parity bits for fault detection. Although this technique is considered cost-effective (e.g., when compared to spatial Redundancy), additional logic is required for parity-bit generation (parity generator) and decoding (parity checker). Further information redundancy-based schemes exist, such as error-correcting codes (ECC), which can detect and correct errors.
    In \cite{gavrilan_2024}, the concept of \textit{impeccable circuits} \cite{8880507} was applied to \textsc{Keccak}, providing an FD mechanism. 
    According to the principles of impeccable circuits \cite{8880507}, coding-scheme-based checkpoints verify the correctness of the underlying computation results and raise an error in case of a mismatch.
    Here, it detects faults up to three bit-flips.
    In \cite{luo2016}, a predictor/compressor approach with parity checks was presented to protect \textsc{Keccak} against faults.
    For the theta layer, parity checks in the two-dimensional slices are calculated. 
    As the parity checks do not involve the three-dimensional characteristics of the \textsc{Keccak}-state, only odd numbers of faulty bits can be detected.
    While fault detection based on impeccable circuits achieves the same level of fault detection, it comes at a high hardware overhead.
    
    \item \textbf{Combined Redundancy Approaches}:
    Based on the main categories, different redundancy techniques can be combined to form a hybrid FD approach.
    In \cite{mestiri2023}, the approach presented in \cite{kahri2017} and \cite{mestiri2021} to provide a spatial- and time-redundancy-based FD mechanism was adapted for \textsc{Keccak}.
    The fault analysis shows a full fault-detection capability for a single bit-flip.
    In \cite{torresalvardo2022}, an error-correction mechanism for \textsc{Keccak} based on Hamming codes and triple modular redundancy (TMR) was proposed.
    It provides two basic implementations with a pipelined separation and round-based implementation of the mapping functions.
    Here, the registers are protected with a (12,8) Hamming code, thus capable of correcting a single-bit error per register byte. In contrast, the mapping functions are protected with TMR by triplicating each function and applying a voter at the outputs to correct a single fault.
\end{itemize}

Based on the detectability analysis, our FD mechanism can be classified as a low-overhead hybrid spatial- and information-redundancy approach performing (cross-)parity bit checks. It is able to detect up to three-bit flips. 

\section{Conclusion and Future Work}
\label{sec:conclusion}

This paper presented a byte-wise in-place partitioning mechanism of the so-called \textsc{Keccak} state. 
The proposed mechanism is lightweight and serves in all required hash modes. 
This allows us to build a unified \textsc{Sha-3}/\textsc{Shake} engine, covering all standard hash configurations, namely: \textsc{Sha-3-224}, \textsc{Sha-3-256}, \textsc{Sha-3-384}, \textsc{Sha-3-512}, \textsc{Shake-128}, and \textsc{Shake-256}.
Thus, the proposed engine suits modern cryptography and standard post-quantum cryptography (PQC) schemes.
Such a unified engine design also meets the requirements of lightweight cryptography, specifically in terms of area and power.

Then, we proposed a lightweight fault-detection mechanism protecting the \textsc{Keccak} state from bit-flip-induced faults.
Here, we introduced a dedicated \textsc{Keccak} register-state protection to achieve sufficient protection. 
We exploited the \textsc{Keccak} state cube and functional characteristics to apply a lightweight cross-parity check mechanism relying on a combined spatial- and information redundancy scheme.
Based on an existing column-sum calculation performed on the \textsc{Keccak} state called \textit{c-plane}, we added a lane-based sum called \textit{f-slice}.
In conjunction with a redundancy module, the combined sums provide a two-dimensional cross-parity check along a sheet, denoted as \textit{z-sheet}.
The evaluation of fault detectability shows that our proposed solution is efficient against up to three-bit flips.
Further, we integrated the lightweight fault detection into an accordingly lightweight unified hash engine for both \textsc{Sha-3} and \textsc{Shake}.

For demonstration and evaluation, we synthesized our design for the 45\,nm FreePDK45 standard library and a Xilinx Virtex-7 FPGA.
The added \textit{z-sheet}-based protection module shows a vast reduction in area overhead compared to state-of-the-art solutions. With only 56\% area overhead compared to 200\%, it shows a 3.7$\times$ improvement. Our resulting \textit{z-sheet}-protected hash engine demonstrates an even higher 4.5$\times$ area-overhead improvement compared to the state of the art.
We furthermore integrated the fault-resilient hash engine into the PULPissimo SoC using their FPGA-centric design environment to showcase the usage of our design in resource-constrained microcontroller-based applications:
On a Xilinx Virtex-7 FPGA, the engine increases the overall PULPissimo SoC by less than 8\,\%.
These results emphasize the lightweightness of the proposed protection mechanism, proving it suitable for microcontroller-based and embedded IoT applications.

In our future work,  we will address lightweight protection of \textsc{Keccak}'s logic layer beyond the state of the art. 
This is required to increase the unified engine's reliability and security against faulting an arbitrary number of bits within a larger chip area.

\received{20 February 2007}
\received[revised]{12 March 2009}
\received[accepted]{5 June 2009}

\begin{acks}
This work was partially funded by the German Ministry of Education and Research (BMBF) via project RILKOSAN (16KISR010K).
\end{acks}

\bibliographystyle{ACM-Reference-Format}
\bibliography{bibliography.bib}

\appendix

\end{document}